\documentclass[nofootinbib,preprint, prX]{revtex4-1} 

\usepackage{amsmath}
\usepackage{mathrsfs}
\usepackage{amssymb}		
\usepackage{graphicx}		
\usepackage{graphics}
\usepackage[small,compact]{titlesec}
\usepackage{xspace}
\usepackage{natbib}
\usepackage{colordvi}
\usepackage{units}
\usepackage{hyperref}
\usepackage{stackrel}
\usepackage{subfigure}
\usepackage{float}

\newcommand{\vect}[1]{\vec{\mathbf{#1}}}
\newcommand{\vectS}[1]{\vec{\boldsymbol{#1}}}

\newcommand{\gTSOTPO}{g^{({}^{3}\!S_{1}-{}^{3}\!P_{1})}}
\newcommand{\gTSOOPO}{g^{({}^{3}\!S_{1}-{}^{1}\!P_{1})}}
\newcommand{\gDIZ}{g^{({}^{1}\!S_{0}-{}^{3}\!P_{0})}_{(\Delta I=0)}}
\newcommand{\gDIO}{g^{({}^{1}\!S_{0}-{}^{3}\!P_{0})}_{(\Delta I=1)}}
\newcommand{\gDIT}{g^{({}^{1}\!S_{0}-{}^{3}\!P_{0})}_{(\Delta I=2)}}

\newcommand{\EFT}{$\mathrm{EFT}(/\!\!\!\pi)$\xspace}
\newcommand{\CG}[6]{C_{#1,#2,#3}^{#4,#5,#6}}
\newcommand{\SJ}[6]{\left\{\begin{array}{ccc} #1 & #2 & #3 \\ #4 & #5 & #6\end{array}\right\}}

\newcommand{\comment}[1]{}

\newcommand{\QS}{{}^{4}\!S_{\nicefrac{3}{2}}}
\newcommand{\DS}{{}^{2}\!S_{\nicefrac{1}{2}}}
\newcommand{\DPoh}{{}^{2}\!P_{\nicefrac{1}{2}}}

\newcommand{\QPoh}{{}^{4}\!P_{\nicefrac{1}{2}}}

\newcommand{\NN}{N\!N}
\newcommand{\Scalo}{\mathcal{S}_{1}}
\newcommand{\Scalt}{\mathcal{S}_{2}}
\newcommand{\Tcal}{\mathcal{T}}

\makeatletter

\newcommand{\Rmnum}[1]{\expandafter\@slowromancap\romannumeral #1@}
\makeatother

\begin{document}

\title{Parity-Violating Three Nucleon Interactions at Low Energies and Large-$N_{C}$}

\author{Jared Vanasse}
\email{jvanasse@stetson.edu}
\affiliation{Department of Physics,
Stetson University,
DeLand, FL 32723
}

\date{\today}

\begin{abstract}
The parity-violating (PV) nucleon-nucleon ($\NN$) interaction in the three-nucleon system is investigated using pionless effective field theory (\EFT).  This work shows that a next-to-leading order (NLO) PV three-body force is necessary in contradiction with a previous claim~\cite{Griesshammer:2010nd}.  Including three-body $P$ to $D$-wave transitions PV three-nucleon observables are calculated to higher energies than previously considered.  Using the recent large-$N_{C}$ analysis of the PV $\NN$ interaction in \EFT the current understanding of low energy PV few-body measurements is reassessed.  The recent measurement of the asymmetry $A_{\gamma}$ in $\vec{n}p\to d\gamma$ from the NPDGamma collaboration~\cite{Blyth:2018aon} gives the value $g_{4}^{(N_{C}^{-1})}=(-1.4\pm 0.63(stat.)\pm 0.09(syst.))\times 10^{-10}~\mathrm{MeV}^{-1}$ for a next-to-next-to-leading order (N$^{2}$LO) in large-$N_{C}$ low energy constant (LEC).  Using the large-$N_{C}$ hierarchy of LECs the sizes of the leading order (LO) in large-$N_{C}$ LECs are estimated  using an experimental bound on a parity violating asymmetry in $\vec{p}d$ scattering at $E_{\mathrm{lab}}=15$~MeV and a measurement of $\vec{pp}$ scattering at $E_{\mathrm{lab}}=13.6$~MeV.  Comparing the size of the resulting LO($\mathcal{O}(N_{C})$) in large-$N_{C}$ LECs to the N$^{2}$LO($\mathcal{O}(N_{C}^{-1})$) in large-$N_{C}$ LEC $g_{4}^{(N_{C}^{-1})}$ shows they are roughly the same size in contradiction with current large-$N_{C}$ counting.

\end{abstract}


\maketitle

\section{Introduction}

Hadronic parity-violation in the Standard Model arises from the exchange of W and Z bosons between quarks, which below $\sim$100~GeV can be described by an effective four quark interaction.  Although the charged-current non-leptonic weak interaction has been studied extensively through hadronic decays the neutral-current non-leptonic weak interaction has not, because its contributions are suppressed compared to the charged-current in hadronic decays.  Hadronic parity-violation offers a probe to study the neutral-current non-leptonic weak current in the Standard Model because at the quark level the $\Delta I=1$ contribution from the charged-current-current four quark interaction is Cabibbo suppressed by $\tan^{2}\theta_{C}\sim 0.04$ compared to the neutral-current-current four quark interaction.  At energies below $\Lambda_{\mathrm{QCD}}$ this four quark interaction is dressed by a complex exchange of gluons and virtual quarks arising from the nonperturbative nature of QCD whose solution is a nontrivial task.  Lattice QCD offers an avenue to solve this~\cite{Beane:2002ca,Wasem:2011zz,Kurth:2015cvl}.  Thus hadronic parity violation offers a unique probe of both fundamental weak and strong physics of the Standard Model.

In nuclear systems, hadronic parity-violation has traditionally been investigated in terms of the DDH (Desplanques, Donoghue, and Holstein) model~\cite{Desplanques:1979hn}.  It consists of seven phenomenological parity-violating (PV) couplings between nucleons arising from the exchange of pseudoscalar and vector mesons.  DDH estimated reasonable ranges and best guesses for the values of these couplings using a quark model and SU(6)$_{W}$ symmetry~\cite{Lipkin:1965ij}.  A more modern approach to describe hadronic parity violation in nuclear systems is provided by effective field theory (EFT), which is model independent and systematically improvable~\cite{Zhu:2004vw,Schindler:2013yua}.  At low energies in EFT the PV nucleon-nucleon ($\NN$) interaction is characterized by five unknown low energy constants (LECs)~\cite{DANILOV196540,Girlanda:2008ts}.  These LECs must be determined from experiment or calculated at the quark level from the fundamental PV effective four quark interactions using Lattice QCD~\cite{Wasem:2011zz}.  In order to cleanly extract the LECs from experiment observables for few-body nuclear systems should be measured for which reliable theoretical calculations can be made.  Recently, a large-$N_{C}$ analysis has shown that not all of the LECs are equally significant~\cite{Phillips:2014kna,Schindler:2015nga}.  In the large-$N_{C}$ counting~\cite{Phillips:2014kna,Schindler:2015nga,Gardner:2017xyl} a linear combination of the isoscalar LECs and the isotensor LEC are  leading-order (LO) or $\mathcal{O}(N_{C})$ in large-$N_C$ counting, while  another linear combination of the isoscalar LECs and the isovector LECs are suppressed by $\sim\!\!1/N_{C}^{2}$.  Thus at LO($\mathcal{O}(N_{C})$) in large-$N_{C}$ PV $\NN$ interactions are characterized by two LECs and this has been shown to be consistent with available experimental data~\cite{Gardner:2017xyl}.  Note, Ref.~\cite{Gardner:2017xyl} did not consider recent results from the NPDGamma collaboration~\cite{Blyth:2018aon} as it was unavailable at the time.  

A self consistent theoretical framework to combine parity-conserving (PC) and PV interactions at low energies ($E<m_{\pi}^{2}/M_{N}$) is provided by pionless effective field theory (\EFT).  The power counting of \EFT, in powers of $Q/\Lambda_{\not{\pi}}$ makes it systematically improvable and allows for estimation of theoretical errors, where $Q$ is a typical momentum scale and $\Lambda_{\not{\pi}}\!\sim\! m_{\pi}$.  Unlike its higher energy cousin chiral-EFT, the power counting of \EFT is well understood and unambiguous~\cite{Valderrama:2016koj}.  \EFT has had great success in describing PC (See Refs.~\cite{Beane:2000fx,Vanasse:2016jtc} for a review) and PV (See Ref.~\cite{Schindler:2013yua} for a review) properties of few-nucleon systems.

PV asymmetries of few nucleon systems are roughly of the size $G_{F}m_{\pi}^{2}\sim 10^{-7}$ and require precision experiments.  At energies where \EFT is valid there is currently only three trusted few-body nonzero PV measurements, two of the longitudinal asymmetry in $pp$ scattering at lab energies of~\cite{Eversheim:1991tg,Nagle:1978vn}
\begin{equation}
A_{L}^{\vec{pp}}=\left\{\begin{array}{rr}
(-0.93\pm 0.20\pm 0.05)\times 10^{-7} & 13.6~\mathrm{MeV}\\
(-1.7\pm 0.8)\times 10^{-7} & 15~\mathrm{MeV}
\end{array}\right.,
\end{equation}
and the photon asymmetry $A_{\gamma}$ in $\vec{n}p\to d\gamma$ from the NPDGamma collaboration~\cite{Blyth:2018aon} 
\begin{equation}
\label{eq:NPDGamma}
A_{\gamma}=(-3.0\pm 1.4(stat.)\pm 0.2(sys.))\times 10^{-8}.
\end{equation}
Both $A_{\gamma}$~\cite{Schindler:2009wd} and $A_{L}^{\vec{pp}}$~\cite{Phillips:2008hn} have been calculated in \EFT.  While the asymmetry from $\vec{p}p$ scattering is sensitive to the two LO LECs in the large-$N_{C}$ counting, $A_{\gamma}$ is primarily sensitive to the isovector contribution in the ${}^{3}\!S_{1}\!\to\!{}^{3}\!P_{1}$ channel, which is next-to-next-to-leading-order (N$^{2}$LO) in the large-$N_{C}$ counting~\cite{Gardner:2017xyl}.  Another possible few-body PV experiment is the asymmetry in the photodisintegration cross-section of deuterium using circularly polarized photons, $P_{\gamma}$, in the process $d\vec{\gamma}\to np$, which could be carried out at an upgraded High Intensity Gamma Ray Source (HI$\gamma$S)	at the Triangle Universities Nuclear Laboratory~\cite{Ahmed:2013jma} (TUNL).  This experiment has been calculated at threshold in \EFT~\cite{Shin:2009hi,Schindler:2009wd} and the ideal energy at which to run it was considered in Ref.~\cite{Vanasse:2014sva}.  It has the advantage of being sensitive to the two LO($\mathcal{O}(N_{C})$) LECs in the large $N_{C}$ counting.  Thus $d\vec{\gamma}\to np$ in combination with $\vec{p}p$ scattering would completely characterize the LO($\mathcal{O}(N_{C})$) in large-$N_{C}$ behavior of the two-body PV LECs.

Three-nucleon measurements offer another potential avenue to study PV interactions.  The PV isotensor contribution in three-nucleon systems is highly suppressed since $\Delta I=2$ cannot connect the isospin-1/2 $N\!d$ system to itself without isospin violation in the PC sector.  Hence, the three-nucleon sector is only sensitive to one of the LO($\mathcal{O}(N_{C})$) LECs in large-$N_{C}$ counting.  Three-nucleon experiments have measured a bound for the longitudinal asymmetry in $\vec{p}d$ scattering at 15~MeV~\cite{Knyazkov:1984lzj} and a $\gamma$-ray asymmetry in the capture of polarized neutrons in $\vec{n}d\to t\gamma$~\cite{Avenier:1984is}.  However, the latter measurement is much larger than expected likely due to an unidentified experimental systematic.  Meanwhile, theoretical \EFT calculations have investigated the spin rotation of a neutron through deuterium~\cite{Griesshammer:2011md,Vanasse:2011nd} and the longitudinal asymmetries in $N\!d$ scattering~\cite{Vanasse:2011nd}. In this work longitudinal asymmetries in $N\!d$ scattering are calculated to higher energies than Ref.~\cite{Vanasse:2011nd} and include three-body $P$ to $D$-wave contributions.  Implications of large-$N_{C}$ on two and three-nucleon PV experiments are also considered.

Another important matter in the three-nucleon system is the order at which three-body PV forces first occur.  Grie{\ss}hammer and Schindler demonstrated that up to and including next-to-leading-order (NLO) in \EFT there is no PV three-body force~\cite{Griesshammer:2010nd}.  This implies that two and three-body PV experiments can be described to $\sim$10\% accuracy with only five two-body PV LECs.  Assuming there are no significant higher body PV forces this should also hold for $A>3$.  The argument made by Grie{\ss}hammer and Schindler for the nonexistence of a NLO PV three-body force relied on Fierz rearrangements.  However, as will be shown the spin-isospin structure on which their argument relies is not invariant under Fierz rearrangements.  In addition, it will be shown by a rigorous asymptotic analysis that a PV three-body force is needed at NLO, and that there is no need for a LO PV three-body force in agreement with Grie{\ss}hammer and Schindler~\cite{Griesshammer:2010nd}.

This paper is organized as follows. Section~\ref{sec:lagr} gives the necessary Lagrangians and discusses two-nucleon scattering.  In Section~\ref{sec:scatt} LO PC and PV $N\!d$ scattering is described and in Section~\ref{sec:asymp} their asymptotic forms are given.  Section~\ref{sec:NLO} reviews the arguments by Grie{\ss}hammer and Schindler for the nonexistence of a NLO PV three-body force and then performing an asymptotic analysis of the NLO PV scattering amplitude demonstrates the necessity for a NLO PV three-body force.  In Section~\ref{sec:observables} PV three-nucleon observables are calculated and the consequences of large-$N_{C}$ on few-body PV measurements is discussed.  Finally, conclusions are given in Section \ref{sec:conclusion}.

\newpage
\section{\label{sec:lagr}Lagrangian}

The NLO PC Lagrangian including two and three-body terms in \EFT is given by 
\begin{align}
&\mathcal{L}=\hat{N}^{\dagger}\left(i\partial_{0}+\frac{\vect{\nabla}^{2}}{2M_{N}}\right)\hat{N}+\hat{t}_{i}^{\dagger}\left[\Delta_{t}-c_{0t}\left(i\partial_{0}+\frac{\vect{\nabla}^{2}}{4M_{N}}+\frac{\gamma_{t}^{2}}{M_{N}}\right)\right]\hat{t}_{i}\\\nonumber
&+\hat{s}_{a}^{\dagger}\left[\Delta_{s}-c_{0s}\left(i\partial_{0}+\frac{\vect{\nabla}^{2}}{4M_{N}}+\frac{\gamma_{s}^{2}}{M_{N}}\right)\right]\hat{s}_{a}-y\left[\hat{t}_{i}^{\dagger}\hat{N} ^{T}P_{i}\hat{N}+\hat{s}_{a}^{\dagger}\hat{N}^{T}\bar{P}_{a}\hat{N}+\mathrm{H.c.}\right]\\\nonumber
&+\frac{y^{2}M_{N}(H_{\mathrm{LO}}(\Lambda)+H_{\mathrm{NLO}}(\Lambda))}{3\Lambda^{2}}\left[\hat{t}_{i}(\sigma_{i}\hat{N})-\hat{s}_{a}(\tau_{A}\hat{N})\right]^{\dagger}\left[\hat{t}_{i}(\sigma_{i}\hat{N})-\hat{s}_{a}(\tau_{A}\hat{N})\right],
\end{align}
where $\hat{N}$ is the nucleon field and $\hat{t}_{i}$ ($\hat{s}_{a}$) is the spin-triplet (spin-singlet) dibaryon field.  The projector $P_{i}=\frac{1}{\sqrt{8}}\sigma_{2}\sigma_{i}\tau_{2}$ ($\bar{P}_{a}=\frac{1}{\sqrt{8}}\sigma_{2}\tau_{2}\tau_{a}$) projects out the spin-triplet iso-singlet (spin-singlet iso-triplet) combination of nuclei.  Using the $Z$-parametrization~\cite{Phillips:1999hh,Griesshammer:2004pe} the two-body parameters are fit to reproduce the poles in the $^{3}\!S_{1}$ and $^{1}\!S_{0}$ channels at LO and their residues about the poles at NLO.  This leads to the values~\cite{Griesshammer:2004pe}
\begin{align}
&y^{2}=\frac{4\pi}{M_{N}},\quad \Delta_{t}=\gamma_{t}-\mu,\quad c_{0t}^{(n)}=(-1)^{n}(Z_{t}-1)^{n+1}\frac{M_{N}}{2\gamma_{t}},\\\nonumber
&\phantom{y_{s}^{2}=\frac{4\pi}{M_{N}},}\quad \Delta_{s}=\gamma_{s}-\mu,\quad \!c_{0s}^{(n)}=(-1)^{n}(Z_{s}-1)^{n+1}\frac{M_{N}}{2\gamma_{s}},
\end{align}
where $\gamma_{t}=45.7025$~MeV ($\gamma_{s}=-7.890$~MeV) is the deuteron binding momentum ($^{1}\!S_{0}$ virtual bound state binding momentum)and $Z_{t}=1.6908$ ($Z_{s}=0.9015$) is the residue about the $^{3}\!S_{1}$ ($^{1}\!S_{0}$) pole.  The value $\mu$ is a mass scale arising from power divergence subtraction~\cite{Kaplan:1998tg} with dimensional regularization.  Note, all physical observables are independent of $\mu$.  The LO and NLO three-body force, $H_{\mathrm{LO}}(\Lambda)$ and $H_{\mathrm{NLO}}(\Lambda)$ respectively, are fit~\cite{Vanasse:2015fph} to the doublet $S$-wave $nd$ scattering length $a_{nd}=0.65$~fm~\cite{Dilg:1971pl}.  The scale $\Lambda$ comes from using cutoff regularization in three-body calculations.

The LO $\NN$ scattering amplitude is given by an infinite sum of diagrams\cite{Kaplan:1998tg,Kaplan:1998we}, which can be solved via a geometric series leading to
\begin{equation}
i\mathcal{A}_{\{t,s\}}=\frac{4\pi}{M_{N}}D_{\{t,s\}}(E,0),
\end{equation}
in the center-of-mass (c.m.) frame where
\begin{equation}
D_{\{t,s\}}(E,q)=\frac{1}{\sqrt{\frac{3}{4}q^{2}-M_{N}E-i\epsilon}-\gamma_{\{t,s\}}},
\end{equation}
is the LO dibaryon propagator with the subscript $t$ ($s$) representing the ${}^{3}\!S_{1}$ (${}^{1}\!S_{0}$) channel.  Taking the residue about the pole of the spin-triplet dibaryon propagator gives the LO deuteron wavefunction renormalization
\begin{equation}
\label{eq:renorm}
Z_{\mathrm{LO}}=\frac{2\gamma_{t}}{M_{N}}.
\end{equation}

Low energy $\NN$ parity-violation is characterized by five independent two-body LEC's described in \EFT by the Lagrangian~\cite{Schindler:2009wd}
\begin{align}
\mathcal{L}_{\mathrm{PV}}=-&\left[\gTSOOPO\hat{t}_{i}^{\dagger}\left(\hat{N}^{T}\sigma_{2}\tau_{2}i\stackrel{\leftrightarrow}{\nabla}_{i}\hat{N}\right)\right.\\\nonumber
&+\gDIZ \hat{s}_{a}^{\dagger}\left(\hat{N}^{T}\sigma_{2}\vectS{\sigma}\cdot\tau_{2}\tau_{a}i\stackrel{\leftrightarrow}{\nabla}\hat{N}\right)\\\nonumber
&+\gDIO\epsilon^{3ab}\hat{s}_{a}^{\dagger}\left(\hat{N}^{T}\sigma_{2}\vectS{\sigma}\cdot\tau_{2}\tau_{b}\stackrel{\leftrightarrow}{\nabla}\hat{N}\right)\\\nonumber
&+\gDIT\mathcal{I}^{ab}\hat{s}_{a}^{\dagger}\left(\hat{N}^{T}\sigma_{2}\vectS{\sigma}\cdot\tau_{2}\tau_{b}i\stackrel{\leftrightarrow}{\nabla}\hat{N}\right)\\\nonumber
&\left.+\gTSOTPO\epsilon^{ijk}\hat{t}_{i}^{\dagger}\left(\hat{N}^{T}\sigma_{2}\sigma_{k}\tau_{2}\tau_{3}\stackrel{\leftrightarrow}{\nabla}_{j}\hat{N}\right)\right]+\mathrm{H.c.},
\end{align}
where $\stackrel{\leftrightarrow}{\nabla}=\stackrel{\rightarrow}{\nabla}-\stackrel{\leftarrow}{\nabla}$ and $\mathcal{I}^{ab}=\mathrm{diag}(1,1,-2)$.  Following Ref.~\cite{Vanasse:2011nd} the definitions
\begin{equation}
g_{1}=\frac{\gTSOOPO}{y},g_{2}=\frac{\gTSOTPO}{y},g_{3}=\frac{\gDIZ}{y},g_{4}=\frac{\gDIO}{y},g_{5}=\frac{\gDIT}{y},
\end{equation}
are made for the two-body PV LECs, which helps to simplify expressions.  Using these definitions the linear combinations
\begin{align}
\label{eq:LECLargeNc}
&g_{1}^{(N_{C})}=\frac{1}{4}g_{1}+\frac{3}{4}g_{3}\,\,\, ,\,\,\,g_{2}^{(N_{C})}=g_{5} && \mathrm{LO}\\\nonumber
g_{3}^{(N_{C}^{-1})}=&\frac{1}{4}g_{1}-\frac{3}{4}g_{3}\,\,\,,\,\,\,g_{4}^{(N_{C}^{-1})}=g_{2}\,\,\,,\,\,\,g_{5}^{(N_{C}^{-1})}=g_{4}&& \mathrm{N}^{2}\mathrm{LO},
\end{align}
of the two-body PV LECs are defined, in which the first line are LECs LO in large-$N_{C}$ and the second line LECs suppressed by $\sim1/N_{C}^{2}$~\cite{Gardner:2017xyl}.  The actual scaling of each LEC is~\cite{Schindler:2015nga} 
\begin{align}
&g_{1}^{(N_{C})}\sim N_{C}\,\,\, ,\,\,\,g_{2}^{(N_{C})}\sim N_{C}\sin^{2}\theta_W && \mathrm{LO}(\mathcal{O}(N_{C}))\\\nonumber
g_{3}^{(N_{C}^{-1})}\sim &N_{C}^{-1}\,\,\,,\,\,\,g_{4}^{(N_{C}^{-1})}\sim N_{C}^{0}\sin^{2}\theta_W\,\,\,,\,\,\,g_{5}^{(N_{C}^{-1})}\sim N_{C}^{0}\sin^{2}\theta_W&& \mathrm{N}^{2}\mathrm{LO}(\mathcal{O}(N_{C}^{-1})),
\end{align}
where factors of $\sin^{2}\theta_W\sim 0.24$ are treated as $1/N_{C}$ corrections.  Note, that Refs.~\cite{Phillips:2014kna,Gardner:2017xyl} did not contain the factor of $\sin^{2}\theta_W$ on the $g_{2}^{(N_{C})}$ LEC as it was only later discovered in Ref.~\cite{Schindler:2015nga}.  Despite $\sin^{2}\theta_W$ appearing to be an additional factor of $N_{C}^{-1}$ on the $g_{2}^{(N_{C})}$ LEC, comparison to experiment in Ref.~\cite{Schindler:2015nga} indicates that $g_{1}^{(N_{C})}$ and $g_{2}^{(N_{C})}$ are of roughly the same size.  This suggests that there is little suppression from $\sin^{2}\theta_W$ at hadronic scales for the isotensor contribution $g_{2}^{(N_{C})}$.  Therefore, in this work $g_{2}^{(N_{C})}$ is kept as LO($\mathcal{O}(N_{C})$) in large-$N_{C}$ counting.

\section{\label{sec:scatt}Leading-Order Scattering}
\subsection{Parity Conserving}

\begin{figure}[hbt]
\includegraphics[width=100mm]{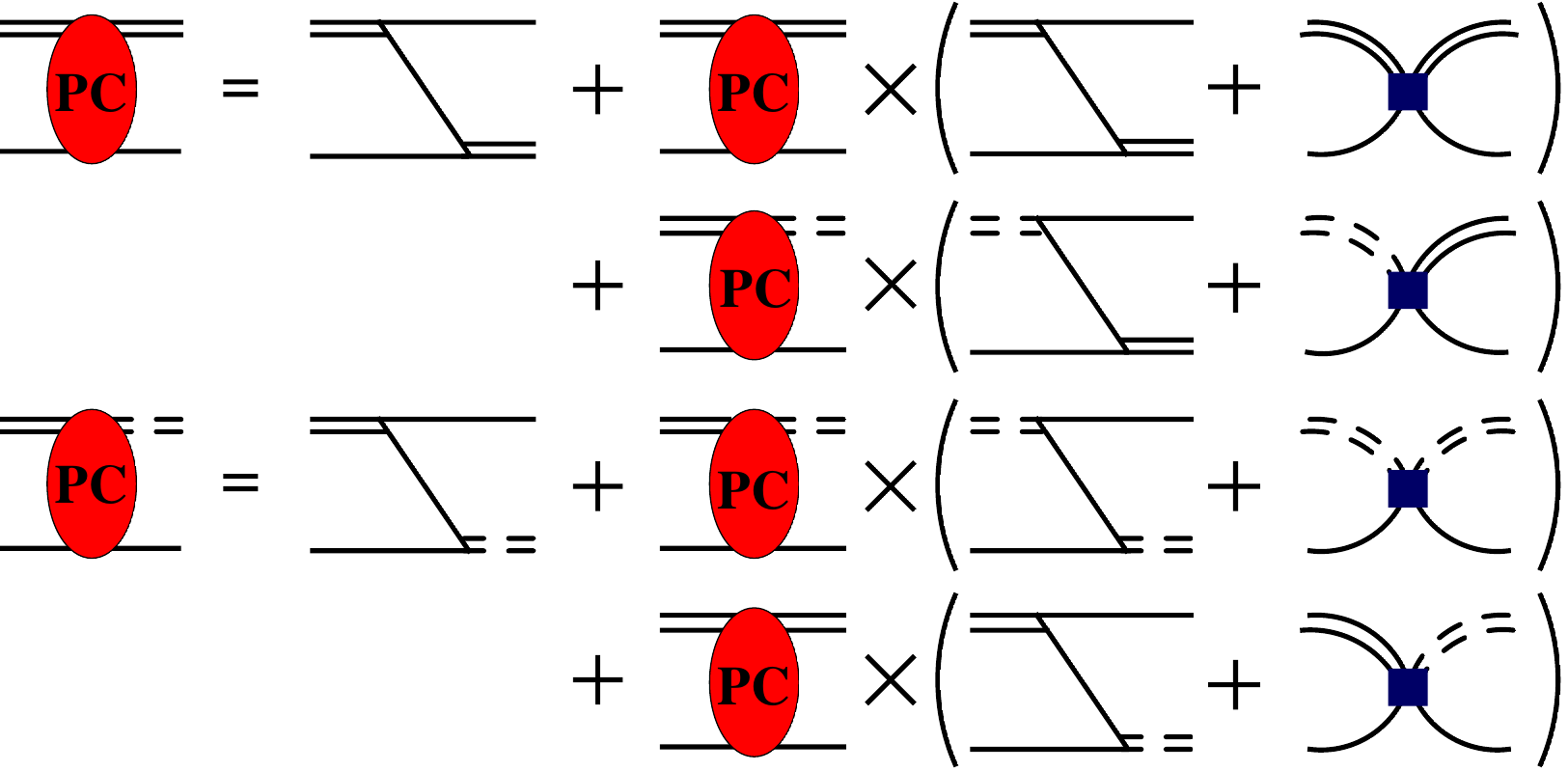}
\caption{\label{fig:LOPC} Coupled integral equations for the LO PC scattering amplitude.  The double (double-dashed) line is the spin-triplet (spin-singlet) dibaryon, the single line a nucleon, the blue box the LO PC three-body force, and the red oval the LO PC scattering amplitude.  Three-body force diagrams only appear in the $\DS$ channel.  In the quartet channel, all diagrams with three-body forces or spin-singlet dibaryons are not present.} 
\end{figure}

The LO half off-shell PC scattering amplitude is given by the integral equation represented in Fig.~\ref{fig:LOPC},
Projecting the integral equation into spin, isospin, and angular momentum the LO half-off shell PC scattering amplitude is given by the integral equation
\begin{align}
\label{eq:PCInegraleq}
{\mathbf{t}_{\mathrm{PC}}}^{J}_{L'S',LS}(k,p,E)=&{\mathbf{K}_{\mathrm{PC}}}^{J}_{L'S',LS}(k,p,E)\mathbf{v}_{p}\\\nonumber
&+\sum_{L'',S''}{\mathbf{K}_{\mathrm{PC}}}^{J}_{L'S',L''S''}(q,p,E)\mathbf{D}\left(E,q\right)\otimes{\mathbf{t}_{\mathrm{PC}}}^{J}_{L''S'',LS}(k,q,E),
\end{align}
where $L$ ($L'$) is the total incoming (outgoing) orbital angular momentum, $S$ ($S'$) is the total incoming (outgoing) spin, and $J$ is the total angular momentum.  $k$ ($p$) is the magnitude of the incoming on-shell (outgoing off-shell) c.m.~momentum and $E=\frac{3k^{2}}{4M_{N}}-\frac{\gamma_{t}^{2}}{M_{N}}$ is the total energy of the $N\!d$ system.  ${\mathbf{t}_{\mathrm{PC}}}^{J}_{L'S',LS}(k,p,E)$ represents a vector in cluster-configuration (c.c.) space~\cite{Griesshammer:2004pe}
\begin{equation}
{\mathbf{t}_{\mathrm{PC}}}^{J}_{L'S',LS}(k,p,E)=\left(\begin{array}{c}
{{t}_{\mathrm{PC}}}^{J;Nt\to Nt}_{L'S',LS}(k,p,E)\\
{{t}_{\mathrm{PC}}}^{J;Nt\to Ns}_{L'S',LS}(k,p,E)
\end{array}\right),
\end{equation}
where ${{t}_{\mathrm{PC}}}^{J;Nt\to Nt}_{L'S',LS}(k,p,E)$ is the $N\!d$ scattering amplitude and ${{t}_{\mathrm{PC}}}^{J;Nt\to Ns}_{L'S',LS}(k,p,E)$ is an unphysical scattering amplitude for $N\!d$ going to a nucleon and spin-singlet dibaryon.  The kernel ${\mathbf{K}_{\mathrm{PC}}}^{J}_{L'S',LS}(k,p,E)$ is a matrix in c.c.~space given by
\begin{align}
\label{eq:PCKernel}
&{\mathbf{K}_{\mathrm{PC}}}^{J}_{L'S',LS}(k,p,E)=\\\nonumber
&\delta_{LL'}\delta_{SS'}(-1)^{L}\left\{
\begin{array}{cc}
\frac{2\pi}{kp}Q_{L}\left(\frac{k^{2}+p^{2}-M_{N}E-i\epsilon}{kp}\right)\left(\begin{array}{rr}
1 & -3\\[-2mm]
-3 & 1
\end{array}\right)+\frac{4\pi H_{\mathrm{LO}}(\Lambda)}{\Lambda^{2}}\delta_{L0}
\left(\begin{array}{rr}
1 & -1\\[-2mm]
-1 & 1
\end{array}\right) & ,S=\nicefrac{1}{2}\\
-\frac{4\pi}{kp}Q_{L}\left(\frac{k^{2}+p^{2}-M_{N}E-i\epsilon}{kp}\right)\left(\begin{array}{rr}
1 & 0\\[-2mm]
0 & 0 
\end{array}\right) & ,S=\nicefrac{3}{2}
\end{array}
\right.,
\end{align}
which matrix multiplies the c.c.~space matrix of dibaryon propagators 
\begin{equation}
\mathbf{D}(E,q)=\left(\begin{array}{cc}
D_{t}(E,q) & 0\\
0 & D_{s}(E,q)
\end{array}\right).
\end{equation}
$Q_{L}(a)$ is a Legendre function of the second kind defined by
\begin{equation}
Q_{L}(a)=\frac{1}{2}\int_{-1}^{1}dx\frac{P_{L}(x)}{x-a},
\end{equation}
where $P_{L}(x)$ are the standard Legendre polynomials and $\mathbf{v}_{p}$ is a vector in c.c.~space
\begin{equation}
\mathbf{v}_{p}=\left(\begin{array}{c}
1\\[-2mm]
0
\end{array}\right),
\end{equation}
which picks out the contributions from the kernel where the initial dibaryon propagators are spin-triplet.  The ``$\otimes$'' notation is a shorthand for integration
\begin{equation}
A(q)\otimes B(q)=\frac{1}{2\pi^{2}}\int_{0}^{\Lambda}dqq^{2}A(q)B(q),
\end{equation}
with $\Lambda$ a cutoff used to regulate potential divergences.

\subsection{Parity Violating }

\begin{figure}[hbt]
\includegraphics[width=100mm]{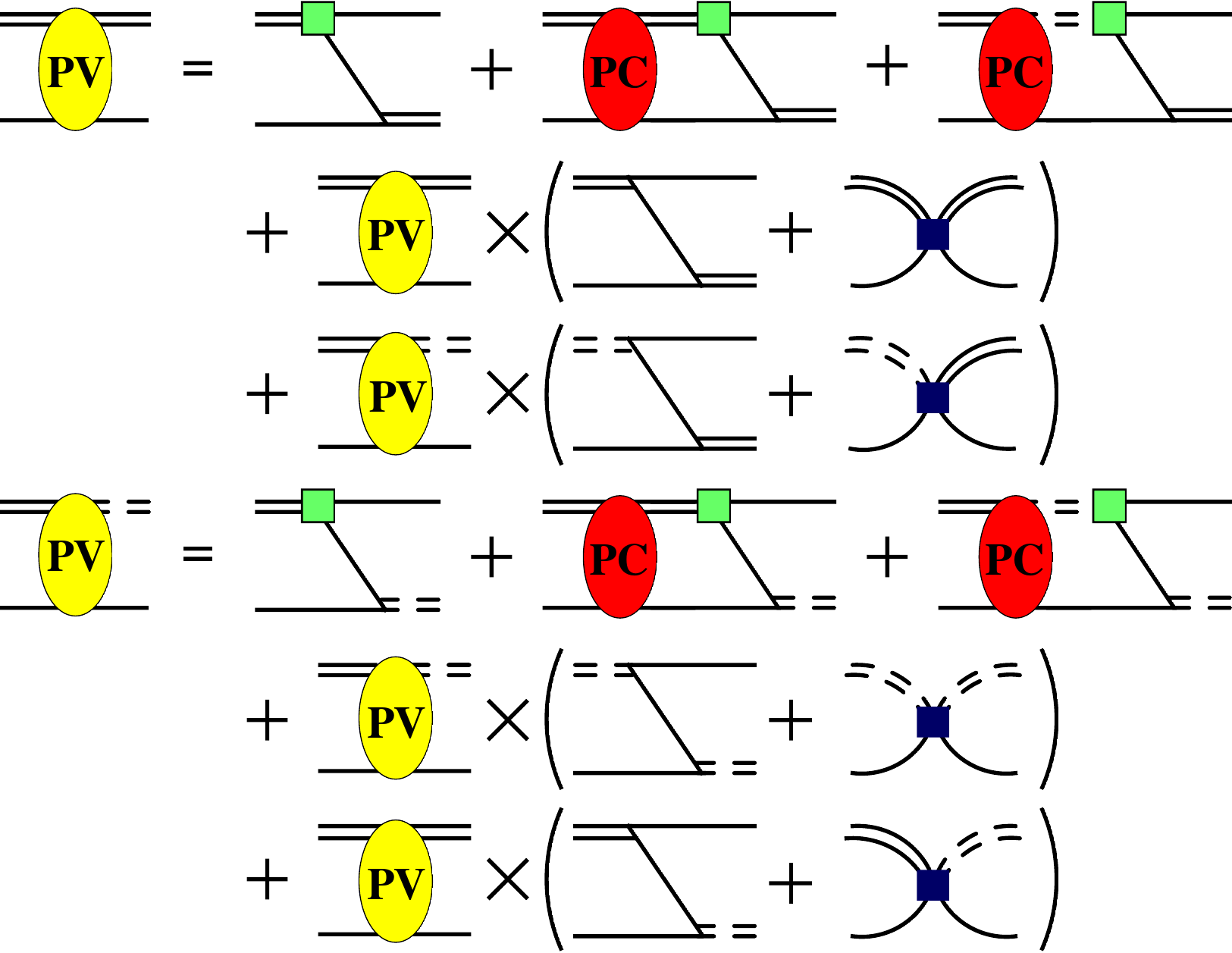}
\caption{\label{fig:PVIntegralEq}Coupled integral equations for the LO PV scattering amplitude.  The green box represents an insertion of a two-body PV LEC and the yellow oval the LO PV scattering amplitude.  Some diagrams only exist in certain channels with the LO three-body force only existing if the outgoing partial wave is $\DS$.} 
\end{figure}

\begin{figure}[hbt]
\includegraphics[width=50mm]{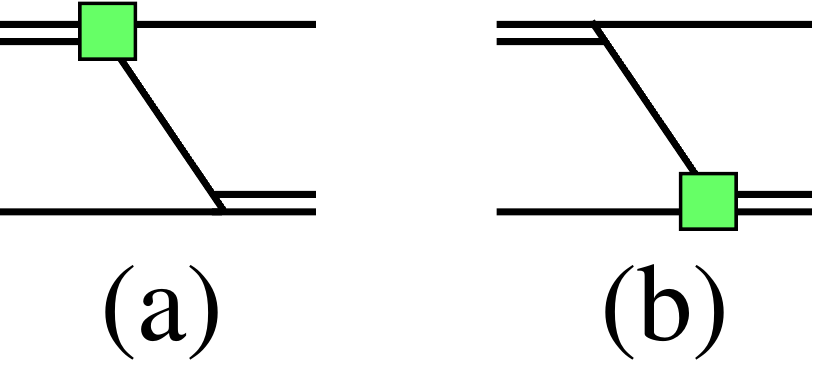}
\caption{\label{fig:treelevel} LO tree-level three-body PV diagrams.  Diagrams (a) and (b) are related by time reversal symmetry and the green box is an insertion of a two-body PV LEC.} 
\end{figure}

The LO PV $N\!d$ scattering amplitude has been calculated previously by convoluting the appropriate LO PC scattering amplitudes with the tree-level PV diagrams in Fig.~\ref{fig:treelevel}~\cite{Griesshammer:2011md,Vanasse:2011nd}.  However, this method does not allow for a straightforward asymptotic analysis of the LO PV scattering amplitude.  Instead, the LO PV scattering amplitude can be represented by the integral equation in Fig.~\ref{fig:PVIntegralEq}~\cite{Vanasse:2011nd},
where diagrams related by time reversal symmetry are not shown.  Projecting in spin, isospin, and angular momentum the diagrams in Fig.~\ref{fig:PVIntegralEq} yield the integral equation
\begin{align}
\label{eq:PVintegralEq}
&{\mathbf{t}_{\mathrm{PV}}}^{J}_{L'S',LS}(k,p,E)={\mathbf{K}_{\mathrm{PV}}}^{J}_{L'S',LS}(k,p,E)\mathbf{v}_{p}\\\nonumber
&\hspace{1cm}+\sum_{L'',S''}{\mathbf{K}_{\mathrm{PV}}}^{J}_{L'S',L''S''}(q,p,E)\otimes\mathbf{D}\left(E,q\right){\mathbf{t}_{\mathrm{PC}}}^{J}_{L''S'',LS}(k,q,E)\\\nonumber
&\hspace{1cm}+\sum_{L'',S''}{\mathbf{K}_{\mathrm{PC}}}^{J}_{L'S',L''S''}(q,p,E)\otimes\mathbf{D}\left(E,q\right){\mathbf{t}_{\mathrm{PV}}}^{J}_{L''S'',LS}(k,q,E),
\end{align}
where
\begin{equation}
{\mathbf{t}_{\mathrm{PV}}}^{J}_{L'S',LS}(k,p,E)=\left(\begin{array}{c}
{{t}_{\mathrm{PV}}}^{J;Nt\to Nt}_{L'S',LS}(k,p,E)\\
{{t}_{\mathrm{PV}}}^{J;Nt\to Ns}_{L'S',LS}(k,p,E)
\end{array}\right),
\end{equation}
is a vector in c.c.~space.  ${\mathbf{K}_{\mathrm{PV}}}^{J}_{L'S',LS}(k,p,E)$ is the projection of the sum of tree-level diagrams containing the two-body PV LECs shown in Fig.~\ref{fig:treelevel}.
Diagram-(a) of Fig.~\ref{fig:treelevel} appears in Fig.~\ref{fig:PVIntegralEq} and diagram-(b) is the time reversed version of diagram-(a) not shown in Fig.~\ref{fig:PVIntegralEq}.  ${\mathbf{K}_{\mathrm{PV}}}^{J}_{L'S',LS}(k,p,E)$ is given by
\begin{equation}
{\mathbf{K}_{\mathrm{PV}}}^{J}_{L'S',LS}(k,p,E)={\mathbf{K}_{\mathrm{PV}}^{(a)}}^{J}_{L'S',LS}(k,p,E)+{\mathbf{K}_{\mathrm{PV}}^{(b)}}^{J}_{L'S',LS}(k,p,E),
\end{equation}
where ${\mathbf{K}_{\mathrm{PV}}^{(a)}}^{J}_{L'S',LS}(k,p,E)$ (${\mathbf{K}_{\mathrm{PV}}^{(b)}}^{J}_{L'S',LS}(k,p,E)$) is the contribution from diagram-(a) (diagram-(b)) in Fig.~\ref{fig:treelevel}.  By time reversal symmetry, contributions from diagram-(a) and (b) are related by
\begin{align}
{\mathbf{K}_{\mathrm{PV}}^{(b)}}^{J}_{L'S',LS}(k,p,E)=\left[{\mathbf{K}_{\mathrm{PV}}^{(a)}}^{J}_{LS,L'S'}(p,k,E)\right]^{T},
\end{align}
where the superscript ``$T$'' denotes the transpose of the c.c.~space matrix.  The $S$ to $P$-wave transitions for the tree level PV diagrams have been calculated previously using projectors in Ref.~\cite{Griesshammer:2011md}, while the general projections for any combination of orbital and spin angular momentum can be found in Ref.~\cite{Vanasse:2011nd}.  These generic projections are updated in Appendix~\ref{app:Projections} following the methods of Ref.~\cite{Margaryan:2015rzg} and include arbitrary isospin.  Following Ref.~\cite{Griesshammer:2011md} the linear combination of LECs
\begin{equation}
\Scalo=3g_{1}+2\tau_{3}g_{2}\quad,\quad\Scalt=3g_{1}-\tau_{3}g_{2}\quad,\quad\Tcal=3g_{3}+2\tau_{3}g_{4}
\end{equation}
are defined, where $\tau_{3}=1$ ($\tau_{3}=-1$) for the $pd$ ($nd$) system.  This set of LECs simplifies the PV kernels and gives the same form for $nd$ and $pd$ systems.  The $S$ to $P$-wave transitions of the PV kernels for diagram-(a) are~\cite{Griesshammer:2011md}
\begin{align}
{\mathbf{K}_{\mathrm{PV}}^{(a)}}^{\frac{1}{2}}_{1\frac{1}{2},0\frac{1}{2}}(k,p,E)&=\frac{4\pi\sqrt{2}}{3}\frac{1}{kp}(2pQ_{0}(a)-kQ_{1}(a))\left(
\begin{array}{cc}
 -\Scalo & \Tcal \\
 -\Scalo & \Tcal \\
\end{array}\right),
\end{align}
\vspace{-5mm}
\begin{align}
{\mathbf{K}_{\mathrm{PV}}^{(a)}}^{\frac{1}{2}}_{0\frac{1}{2},1\frac{1}{2}}(k,p,E)&=\frac{4\pi\sqrt{2}}{3}\frac{1}{kp}(kQ_{0}(a)-2pQ_{1}(a))\left(
\begin{array}{cc}
 -\Scalo & \Tcal \\
 -\Scalo & \Tcal \\
\end{array}\right),
\end{align}
\vspace{-5mm}
\begin{align}
{\mathbf{K}_{\mathrm{PV}}^{(a)}}^{\frac{1}{2}}_{1\frac{3}{2},0\frac{1}{2}}(k,p,E)&=\frac{16\pi}{3}\frac{1}{kp}(2pQ_{0}(a)-kQ_{1}(a))\left(
\begin{array}{cc}
 \frac{(\Scalo-\Scalt)}{3} & \Tcal \\
 0 & 0 \\
\end{array}\right),
\end{align}
\vspace{-5mm}
\begin{align}
{\mathbf{K}_{\mathrm{PV}}^{(a)}}^{\frac{1}{2}}_{0\frac{1}{2},1\frac{3}{2}}(k,p,E)&=\frac{8\pi}{3}\frac{1}{kp}(kQ_{0}(a)-2pQ_{1}(a))\left(
\begin{array}{cc}
 \Scalt & 0 \\
 \Scalt & 0 \\
\end{array}\right),
\end{align}
\vspace{-5mm}
\begin{align}
{\mathbf{K}_{\mathrm{PV}}^{(a)}}^{\frac{3}{2}}_{1\frac{1}{2},0\frac{3}{2}}(k,p,E)&=-\frac{4\pi\sqrt{2}}{3}\frac{1}{kp}(2pQ_{0}(a)-kQ_{1}(a))\left(
\begin{array}{cc}
 \Scalt & 0 \\
 \Scalt & 0 \\
\end{array}\right),
\end{align}
\vspace{-5mm}
\begin{align}
{\mathbf{K}_{\mathrm{PV}}^{(a)}}^{\frac{3}{2}}_{0\frac{3}{2},1\frac{1}{2}}(k,p,E)&=-\frac{8\pi\sqrt{2}}{3}\frac{1}{kp}(kQ_{0}(a)-2pQ_{1}(a))\left(
\begin{array}{cc}
 \frac{(\Scalo-\Scalt)}{3} & \Tcal \\
 0 & 0 \\
\end{array}\right),
\end{align}
\vspace{-5mm}
\begin{align}
{\mathbf{K}_{\mathrm{PV}}^{(a)}}^{\frac{3}{2}}_{1\frac{3}{2},0\frac{3}{2}}(k,p,E)&=\frac{8\pi\sqrt{10}}{3}\frac{1}{kp}(2pQ_{0}(a)-kQ_{1}(a))\left(
\begin{array}{cc}
 \frac{(\Scalo-\Scalt)}{3} & 0 \\
 0 & 0 \\
\end{array}\right),
\end{align}
and
\begin{align}
{\mathbf{K}_{\mathrm{PV}}^{(a)}}^{\frac{3}{2}}_{0\frac{3}{2},1\frac{3}{2}}(k,p,E)&=\frac{8\pi\sqrt{10}}{3}\frac{1}{kp}(kQ_{0}(a)-2pQ_{1}(a))\left(
\begin{array}{cc}
 \frac{(\Scalo-\Scalt)}{3} & 0 \\
 0 & 0 \\
\end{array}\right).
\end{align}

The $P$ to $D$-wave transition projections are\footnote{The $P$ to $D$ wave transitions were independently calculated using projector methods by Trevor Balint and Harald W. Grie{\ss}hammer, and their calculations agree with the results of this work.}
\begin{align}
{\mathbf{K}_{\mathrm{PV}}^{(a)}}^{\frac{3}{2}}_{2\frac{1}{2},1\frac{1}{2}}(k,p,E)&=\frac{4\pi\sqrt{2}}{3}\frac{1}{kp}(-2pQ_{1}(a)+kQ_{2}(a))\left(
\begin{array}{cc}
 -\Scalo & \Tcal \\
 -\Scalo & \Tcal \\
\end{array}\right),
\end{align}
\vspace{-5mm}
\begin{align}
{\mathbf{K}_{\mathrm{PV}}^{(a)}}^{\frac{3}{2}}_{1\frac{1}{2},2\frac{1}{2}}(k,p,E)&=\frac{4\pi\sqrt{2}}{3}\frac{1}{kp}(-kQ_{1}(a)+2pQ_{2}(a))\left(
\begin{array}{cc}
 -\Scalo & \Tcal \\
 -\Scalo & \Tcal \\
\end{array}\right),
\end{align}
\vspace{-5mm}
\begin{align}
{\mathbf{K}_{\mathrm{PV}}^{(a)}}^{\frac{1}{2}}_{2\frac{3}{2},1\frac{1}{2}}(k,p,E)&=\frac{16\pi}{3}\frac{1}{kp}(-2pQ_{1}(a)+kQ_{2}(a))\left(
\begin{array}{cc}
 \frac{(\Scalo-\Scalt)}{3} & \Tcal \\
 0 & 0 \\
\end{array}\right),
\end{align}
\vspace{-5mm}
\begin{align}
{\mathbf{K}_{\mathrm{PV}}^{(a)}}^{\frac{1}{2}}_{1\frac{1}{2},2\frac{3}{2}}(k,p,E)&=\frac{8\pi}{3}\frac{1}{kp}(-kQ_{1}(a)+2pQ_{2}(a))\left(
\begin{array}{cc}
 \Scalt & 0 \\
 \Scalt & 0 \\
\end{array}\right),
\end{align}
\vspace{-5mm}
\begin{align}
{\mathbf{K}_{\mathrm{PV}}^{(a)}}^{\frac{3}{2}}_{2\frac{1}{2},1\frac{3}{2}}(k,p,E)&=-\sqrt{\frac{2}{5}}\frac{4\pi}{3}\frac{1}{kp}(-2pQ_{1}(a)+kQ_{2}(a))\left(
\begin{array}{cc}
 \Scalt & 0 \\
 \Scalt & 0 \\
\end{array}\right),
\end{align}
\vspace{-5mm}
\begin{align}
{\mathbf{K}_{\mathrm{PV}}^{(a)}}^{\frac{3}{2}}_{1\frac{3}{2},2\frac{1}{2}}(k,p,E)&=-\sqrt{\frac{2}{5}}\frac{8\pi}{3}\frac{1}{kp}(-kQ_{1}(a)+2pQ_{2}(a))\left(
\begin{array}{cc}
 \frac{(\Scalo-\Scalt)}{3} & \Tcal \\
 0 & 0 \\
\end{array}\right),
\end{align}
\vspace{-5mm}
\begin{align}
{\mathbf{K}_{\mathrm{PV}}^{(a)}}^{\frac{1}{2}}_{2\frac{3}{2},1\frac{3}{2}}(k,p,E)&=\frac{8\pi\sqrt{2}}{3}\frac{1}{kp}(-2pQ_{1}(a)+kQ_{2}(a))\left(
\begin{array}{cc}
 \frac{(\Scalo-\Scalt)}{3} & 0 \\
 0 & 0 \\
\end{array}\right),
\end{align}
and
\begin{align}
{\mathbf{K}_{\mathrm{PV}}^{(a)}}^{\frac{1}{2}}_{1\frac{3}{2},2\frac{3}{2}}(k,p,E)&=\frac{8\pi\sqrt{2}}{3}\frac{1}{kp}(-kQ_{1}(a)+2pQ_{2}(a))\left(
\begin{array}{cc}
 \frac{(\Scalo-\Scalt)}{3} & 0 \\
 0 & 0 \\
\end{array}\right),
\end{align}
where
\begin{equation}
\label{eq:a}
a=\frac{k^{2}+p^{2}-M_{N}E-i\epsilon}{kp}.
\end{equation}
Transitions with the same orbital and spin angular momentum but different $J$-values are related by overall constants given by
\begin{align}
\label{eq:Jone}
{\mathbf{K}_{\mathrm{PV}}^{(a)}}^{\frac{3}{2}}_{2\frac{3}{2},1\frac{1}{2}}(k,p,E)&=\frac{1}{\sqrt{2}}{\mathbf{K}_{\mathrm{PV}}^{(a)}}^{\frac{1}{2}}_{2\frac{3}{2},1\frac{1}{2}}(k,p,E),
\end{align}
\vspace{-9mm}
\begin{align}
\label{eq:Jtwo}
{\mathbf{K}_{\mathrm{PV}}^{(a)}}^{\frac{3}{2}}_{1\frac{1}{2}\,2\frac{3}{2}}(k,p,E)&=\frac{1}{\sqrt{2}}{\mathbf{K}_{\mathrm{PV}}^{(a)}}^{\frac{1}{2}}_{1\frac{1}{2},2\frac{3}{2}}(k,p,E),
\end{align}
\vspace{-9mm}
\begin{align}
\label{eq:Jthree}
{\mathbf{K}_{\mathrm{PV}}^{(a)}}^{\frac{5}{2}}_{2\frac{1}{2},1\frac{3}{2}}(k,p,E)&=\sqrt{6}{\mathbf{K}_{\mathrm{PV}}^{(a)}}^{\frac{3}{2}}_{2\frac{1}{2},1\frac{3}{2}}(k,p,E),
\end{align}
\vspace{-9mm}
\begin{align}
\label{eq:Jfour}
{\mathbf{K}_{\mathrm{PV}}^{(a)}}^{\frac{5}{2}}_{1\frac{3}{2},2\frac{1}{2}}(k,p,E)&=\sqrt{6}{\mathbf{K}_{\mathrm{PV}}^{(a)}}^{\frac{3}{2}}_{1\frac{3}{2},2\frac{1}{2}}(k,p,E),
\end{align}
\vspace{-9mm}
\begin{align}
\label{eq:Jfive}
{\mathbf{K}_{\mathrm{PV}}^{(a)}}^{\frac{3}{2}}_{2\frac{3}{2},1\frac{3}{2}}(k,p,E)&=\frac{4}{\sqrt{5}}{\mathbf{K}_{\mathrm{PV}}^{(a)}}^{\frac{1}{2}}_{2\frac{3}{2},1\frac{3}{2}}(k,p,E)=\frac{4}{\sqrt{21}}{\mathbf{K}_{\mathrm{PV}}^{(a)}}^{\frac{5}{2}}_{2\frac{3}{2},1\frac{3}{2}}(k,p,E),
\end{align}
and
\begin{align}
\label{eq:Jsix}
{\mathbf{K}_{\mathrm{PV}}^{(a)}}^{\frac{3}{2}}_{1\frac{3}{2},2\frac{3}{2}}(k,p,E)&=\frac{4}{\sqrt{5}}{\mathbf{K}_{\mathrm{PV}}^{(a)}}^{\frac{1}{2}}_{1\frac{3}{2},2\frac{3}{2}}(k,p,E)=\frac{4}{\sqrt{21}}{\mathbf{K}_{\mathrm{PV}}^{(a)}}^{\frac{5}{2}}_{1\frac{3}{2},2\frac{3}{2}}(k,p,E).
\end{align}

\section{\label{sec:asymp}Leading-Order Asymptotic Behavior}
\subsection{Parity Conserving}
To calculate the asymptotic behavior of the scattering amplitudes it is instructive to transform the scattering amplitudes into the Wigner basis defined by the linear combinations~\cite{Bedaque:1999ve}
\begin{equation}
\left(\begin{array}{c}
{{t}_{\mathrm{PC}}}^{J;W\!s}_{L'S',LS}(k,p,E)\\
{{t}_{\mathrm{PC}}}^{J;W\!as}_{L'S',LS}(k,p,E)
\end{array}\right)
=
\left(\begin{array}{c}
{{t}_{\mathrm{PC}}}^{J;Nt\to Nt}_{L'S',LS}(k,p,E)-{{t}_{\mathrm{PC}}}^{J;Nt\to Ns}_{L'S',LS}(k,p,E)\\
{{t}_{\mathrm{PC}}}^{J;Nt\to Nt}_{L'S',LS}(k,p,E)+{{t}_{\mathrm{PC}}}^{J;Nt\to Ns}_{L'S',LS}(k,p,E)
\end{array}\right),
\end{equation}
where $W\!s$ ($W\!as$) is the Wigner symmetric (Wigner anti-symmetric) combination of amplitudes.  In the Wigner basis, consequences of Wigner $SU(4)$ symmetry~\cite{Wigner:1936dx}, which is a combination of spin and isospin into a single four-vector, become apparent.  The Wigner symmetric (Ws) amplitude does not change sign under the interchange of spin and isospin, whereas the Wigner anti-symmetric (Was) amplitude does.  In the Wigner limit ($\gamma_{t}=\gamma_{s}$) the LO integral equations for the Ws and Was scattering amplitudes decouple making the Ws amplitude equivalent to three-bosons~\cite{Bedaque:1999ve,Bedaque:1998km,Bedaque:1998kg}.  When $\gamma_{t}=\gamma_{s}$ the interactions between neutrons and protons become identical at LO giving a purely symmetric spatial wavefunction like three-bosons.  The spin and isospin part of the wavefunction factors out from the symmetric spatial part since there is no other spatial component to mix with.  Expansion about the Wigner limit leads to good predictions of properties of three-nucleon systems~\cite{Vanasse:2016umz}.

Transformation into the Wigner basis is achieved by the matrix projector
\begin{equation}
{\mathbf{t_{W}}_{\mathrm{PC}}}^{J}_{L'S',LS}(k,p,E)=\left(\begin{array}{rr}
1 & -1\\[-2mm]
1 & 1
\end{array}
\right){\mathbf{t}_{\mathrm{PC}}}^{J}_{L'S',LS}(k,p,E),
\end{equation}
where
\begin{equation}
{\mathbf{t_{W}}_{\mathrm{PC}}}^{J}_{L'S',LS}(k,p,E)=
\left(\begin{array}{c}
{{t}_{\mathrm{PC}}}^{J;W\!s}_{L'S',LS}(k,p,E)\\
{{t}_{\mathrm{PC}}}^{J;W\!as}_{L'S',LS}(k,p,E)
\end{array}\right).
\end{equation}
By repeated use of the identity
\begin{equation}
\left(\begin{array}{rr}
1 & 0\\[-2mm]
0 & 1\\
\end{array}\right)=\frac{1}{2}\left(\begin{array}{rr}
1 & 1\\[-2mm]
-1 & 1
\end{array}\right)\left(\begin{array}{rr}
1 & -1\\[-2mm]
1 & 1
\end{array}\right),
\end{equation}
and the matrix projector for the Wigner basis, the LO PC scattering amplitude Eq.~(\ref{eq:PCInegraleq}) can be written into the Wigner basis giving
\begin{align}
&{\mathbf{t_{W}}_{\mathrm{PC}}}^{J}_{L'S',LS}(k,p,E)={\mathbf{K_{W}}_{\mathrm{PC}}}^{J}_{L'S',LS}(k,p,E)\mathbf{v_{W}}_{p}\\\nonumber
&\hspace{1cm}+\sum_{L'',S''}{\mathbf{K_{W}}_{\mathrm{PC}}}^{J}_{L'S',L''S''}(q,p,E)\otimes\mathbf{D_{W}}\left(E,q\right){\mathbf{t_{W}}_{\mathrm{PC}}}^{J}_{L''S'',LS}(k,q,E).
\end{align}
The matrix $\mathbf{D_{W}}(E,q)$ is the dibaryon matrix in the Wigner basis defined by
\begin{equation}
\mathbf{D_{W}}(E,q)=\frac{1}{2}\left(\begin{array}{rr}
1 & -1\\[-2mm]
1 & 1
\end{array}\right)\mathbf{D}(E,q)\left(\begin{array}{rr}
1 & 1\\[-2mm]
-1 & 1
\end{array}\right)=\left(\begin{array}{ll}
D_{W\!s}(E,q) & D_{W\!as}(E,q)\\
D_{W\!as}(E,q) & D_{W\!s}(E,q)
\end{array}\right),
\end{equation}
where
\begin{equation}
D_{W\!s}(E,q)=\frac{1}{2}\left(D_{t}(E,q)+D_{s}(E,q)\right)\quad,\quad D_{W\!as}(E,q)=\frac{1}{2}\left(D_{t}(E,q)-D_{s}(E,q)\right),
\end{equation}
and ${\mathbf{K_{W}}_{\mathrm{PC}}}^{J}_{L'S',LS}(k,p,E)$ is the kernel in the Wigner basis given by
\begin{equation}
{\mathbf{K_{W}}_{\mathrm{PC}}}^{J}_{L'S',LS}(k,p,E)=\frac{1}{2}\left(\begin{array}{rr}
1 & -1\\[-2mm]
1 & 1
\end{array}\right){\mathbf{K}_{\mathrm{PC}}}^{J}_{L'S',LS}(k,p,E)
\left(\begin{array}{rr}
1 & 1\\[-2mm]
-1 & 1
\end{array}\right).
\end{equation}
Projecting the LO PC kernel Eq.~(\ref{eq:PCKernel}) into the Wigner basis gives
\begin{align}
&{\mathbf{K_{W}}_{\mathrm{PC}}}^{J}_{L'S',LS}(k,p,E)=\\\nonumber
&\delta_{LL'}\delta_{SS'}(-1)^{L}\left\{
\begin{array}{cc}
\frac{2\pi}{kp}Q_{L}\left(\frac{k^{2}+p^{2}-M_{N}E-i\epsilon}{kp}\right)\left(\begin{array}{rr}
4 & 0\\[-2mm]
0 & -2
\end{array}\right)+\frac{4\pi H_{\mathrm{LO}}(\Lambda)}{\Lambda^{2}}\delta_{L0}
\left(\begin{array}{rr}
2 & 0\\[-2mm]
0 & 0
\end{array}\right) & ,S=\nicefrac{1}{2}\\
-\frac{2\pi}{kp}Q_{L}\left(\frac{k^{2}+p^{2}-M_{N}E-i\epsilon}{kp}\right)\left(\begin{array}{rr}
1 & 1\\[-2mm]
1 & 1 
\end{array}\right) & ,S=\nicefrac{3}{2}
\end{array}
\right.,
\end{align}
and $\mathbf{v_{W}}_{p}$ is given by
\begin{equation}
\mathbf{v_{W}}_{p}=\left(\begin{array}{rr}
1 & -1\\[-2mm]
1 & 1
\end{array}\right)\mathbf{v}_{p}=\left(\begin{array}{c}
1 \\[-2mm]
1
\end{array}\right).
\end{equation}

After going to the Wigner basis the integral equation must be expanded in the asymptotic limit $p\sim q\gg k,E,\gamma_{t}$, and $\gamma_{s}$.  In this limit, the dibaryon propagators in the Wigner basis can be expanded yielding
\begin{equation}
\label{eq:DWsasymp}
D_{W\!s}\left(E,q\right)\sim\frac{2}{\sqrt{3}}\frac{1}{q}+\frac{4}{3}\frac{\gamma_{t}+\gamma_{s}}{q^{2}}+\cdots,
\end{equation}
and
\begin{equation}
\label{eq:DWasasymp}
D_{W\!as}\left(E,q\right)\sim\frac{4}{3}\frac{\gamma_{t}-\gamma_{s}}{q^{2}}+\cdots.
\end{equation}
Here the utility of the Wigner basis is apparent, because $D_{W\!as}(E,q)$ is subleading compared to $D_{W\!s}(E,q)$ in the asymptotic limit.  This means that in the leading term of the asymptotic expansion the Ws scattering amplitude decouples from the Was amplitude.  In the asymptotic limit the energy present in the Legendre functions of the second kind must also be expanded out~\cite{Ji:2012nj,ji2012universality}, however, to NLO this is not necessary and the energy dependence can be dropped from the Legendre functions of the second kind.  Further details of calculating the asymptotic form of the LO PC scattering amplitudes can be found in Refs.~\cite{Bedaque:2002yg,Griesshammer:2005ga,Vanasse:2014kxa}, below their results are given.

The asymptotic behavior of the $\DS$ scattering amplitude in the Ws channel takes the form~\cite{Bedaque:2002yg,Vanasse:2014kxa}
\begin{equation}
{t_{\mathrm{PC}}}^{\frac{1}{2};W\!s}_{0\frac{1}{2},0\frac{1}{2}}(k,q,E)=C\frac{\sin\left(s_{0}\ln\left(\frac{q}{\Lambda^{*}}\right)\right)}{q}+C\frac{1}{\sqrt{3}}(\gamma_{t}+\gamma_{s})|B_{-1}|\frac{\sin\left(s_{0}\ln\left(\frac{q}{\Lambda^{*}}\right)+\mathrm{Arg}(B_{-1})\right)}{q^{2}}+\cdots,
\end{equation}
and in the Was channel the form
\begin{equation}
{t_{\mathrm{PC}}}^{\frac{1}{2};W\!as}_{0\frac{1}{2},0\frac{1}{2}}(k,q,E)=-\frac{C}{2\sqrt{3}}(\gamma_{t}-\gamma_{s})|\tilde{B}_{-1}|\frac{\sin\left(s_{0}\ln\left(\frac{q}{\Lambda^{*}}\right)+\mathrm{Arg}(\tilde{B}_{-1})\right)}{q^{2}},
\end{equation}
where
\begin{equation}
B_{-1}=\frac{I(is_{0}-1)}{1-I(is_{0}-1)},
\end{equation}
and
\begin{equation}
\tilde{B}_{-1}=\frac{I(is_{0}-1)}{1+\frac{1}{2}I(is_{0}-1)}.
\end{equation}
The value $s_{0}=1.00624...$ comes from solving the transcendental equation, $I(is_{0})=1$, where
\begin{equation}
\label{eq:I}
I(s)=\frac{4}{\sqrt{3}\pi}\int_{0}^{\infty}dx\ln\left(\frac{x^{2}+x+1}{x^{2}-x+1}\right)x^{s-1}=\frac{8}{\sqrt{3}s}\frac{\sin\left(\frac{\pi s}{6}\right)}{\cos\left(\frac{\pi s}{2}\right)},
\end{equation}
comes from the leading asymptotic behavior of the Ws $\DS$ scattering amplitude~\cite{Bedaque:2002yg}.  $C=0.4315$~MeV$^{-1}$ and $\Lambda^{*}=1.6251$~MeV must be determined by fitting the numerically calculated half off-shell scattering amplitude to the asymptotic form of the scattering amplitude.  The value of $\Lambda^{*}$ depends on the three-body force and the regularization method used, while the value of $C$ depends on the infra-red (IR) physics, where $k=1$~MeV is chosen to calculate the LO half off-shell scattering amplitude for different cutoffs $\Lambda$.

The leading asymptotic behavior of the Was $\DPoh$ scattering amplitude calculated by Grie{\ss}hammer~\cite{Griesshammer:2005ga} and its subleading behavior takes the form
\begin{equation}
{t_{\mathrm{PC}}}^{\frac{1}{2};W\!as}_{1\frac{1}{2},1\frac{1}{2}}(k,q,E)\sim B^{\DPoh}q^{-(s_{1}+1)}+B^{\DPoh}_{-1}q^{-(s_{1}+2)}+\cdots.
\end{equation}
$s_{1}=1.77272...$ is determined by the transcendental equation, $-\frac{1}{2}\mathcal{M}[1,s_{1}]=1$,~\cite{Griesshammer:2005ga} where
\begin{equation}
\label{eq:Mellin}
\mathcal{M}[\ell,s]=\frac{8}{\sqrt{3}\pi}\int_{0}^{\infty}dx Q_{\ell}\left(x+\frac{1}{x}\right)x^{s-1},
\end{equation}
is the Mellin transform of $Q_{\ell}\left(x+\frac{1}{x}\right)$ up to a multiplicative constant chosen such that $\mathcal{M}[0,s]=I(s)$.  The general solution for the Mellin transform of $Q_{\ell}\left(x+\frac{1}{x}\right)$ has been calculated by Grie{\ss}hammer~\cite{Griesshammer:2005ga}. $B^{\DPoh}=1.274$~MeV$^{s_{1}-1}$ is determined by fitting to the numerically calculated half off-shell scattering amplitude for which $k=1$~MeV, while the value of $B^{\DPoh}_{-1}$ and its derivation are given in Appendix~\ref{app:PCasymptotic}.  The Ws $\DPoh$ scattering amplitude is subleading compared to the Was part~\cite{Griesshammer:2005ga} and will not be needed.

The asymptotic scaling of the $\QPoh$ channel has the same leading power law scaling as the Was $\DPoh$ channel~\cite{Griesshammer:2005ga}
\begin{equation}
{t_{\mathrm{PC}}}^{\frac{1}{2};W\!as}_{1\frac{3}{2},1\frac{3}{2}}(k,q,E)\sim B^{\QPoh}q^{-(s_{1}+1)}+B^{\QPoh}_{-1}q^{-(s_{1}+2)}+\cdots.
\end{equation}
The leading coefficient $B^{\QPoh}=1.929$~MeV$^{s_{1}-1}$ is again fit to the numerically calculated half off-shell scattering amplitude for which $k=1$~MeV.  The subleading coefficient, $B^{\QPoh}_{-1}$ and its derivation are given in Appendix~\ref{app:PCasymptotic}.

\subsection{Parity Violating}

To carry out an asymptotic analysis on the LO PV scattering amplitude, Eq.~(\ref{eq:PVintegralEq}) must be rewritten into the Wigner basis yielding\footnote{Further details of the asymptotic analysis of the LO $nd$ scattering amplitude can be found in Refs.~\cite{Bedaque:1998km,Bedaque:1999ve}.  Ref.~\cite{Bedaque:2002yg} discusses the asymptotic analysis beyond LO and Ref.~\cite{ji2012universality} gives further details into an asymptotic analysis of similar three-boson systems.}
\begin{align}
\label{eq:PVLOWig}
&{\mathbf{t_{W}}_{\mathrm{PV}}}^{J}_{L'S',LS}(k,p,E)={\mathbf{K_{W}}_{\mathrm{PV}}}^{J}_{L'S',LS}(k,p,E)\mathbf{v_{W}}_{p}\\\nonumber
&\hspace{1cm}+\sum_{L'',S''}{\mathbf{K_{W}}_{\mathrm{PV}}}^{J}_{L'S',L''S''}(q,p,E)\otimes\mathbf{D_{W}}\left(E,q\right){\mathbf{t_{W}}_{\mathrm{PC}}}^{J}_{L''S'',LS}(k,q,E)\\\nonumber
&\hspace{1cm}+\sum_{L'',S''}{\mathbf{K_{W}}_{\mathrm{PC}}}^{J}_{L'S',L''S''}(q,p,E)\otimes\mathbf{D_{W}}\left(E,q\right){\mathbf{t_{W}}_{\mathrm{PV}}}^{J}_{L''S'',LS}(k,q,E),
\end{align}
where
\begin{equation}
{\mathbf{K_{W}}_{\mathrm{PV}}}^{J}_{L'S',LS}(k,p,E)=\frac{1}{2}\left(\begin{array}{rr}
1 & -1\\[-2mm]
1 & 1
\end{array}\right){\mathbf{K}_{\mathrm{PV}}}^{J}_{L'S',LS}(k,p,E)
\left(\begin{array}{rr}
1 & 1\\[-2mm]
-1 & 1
\end{array}\right).
\end{equation}
is defined analogously to ${\mathbf{K_{W}}_{\mathrm{PC}}}^{J}_{L'S',LS}(k,p,E)$.  The only PV channels that have divergences at NLO contain $\DS$ and their respective kernels in the Wigner basis are given by
\begin{align}
\label{eq:Wbasis1}
&{\mathbf{K_{W}}_{\mathrm{PV}}}^{\frac{1}{2}}_{1\frac{1}{2},0\frac{1}{2}}(k,p,E)=4\pi\sqrt{2} \left[ \left(
\begin{array}{cc}
 0 & -\frac{1}{3}(\Tcal+\Scalo) \\
-\frac{2}{3}(\Tcal+\Scalo)  & \Tcal-\Scalo \\
\end{array}
\right)\frac{Q_{0}(a)}{k}\right.\\\nonumber
&\hspace{7cm}-\left.\left(
\begin{array}{cc}
0 & -\frac{2}{3}(\Tcal+\Scalo) \\
-\frac{1}{3}(\Tcal+\Scalo) & \Tcal-\Scalo \\
\end{array}
\right)\frac{Q_1(a)}{p}\right],
\end{align}
\begin{align}
\label{eq:Wbasis2}
&{\mathbf{K_{W}}_{\mathrm{PV}}}^{\frac{1}{2}}_{0\frac{1}{2},1\frac{1}{2}}(k,p,E)=4\pi\sqrt{2} \left[ \left(
\begin{array}{cc}
 0 & -\frac{2}{3}(\Tcal+\Scalo) \\
 -\frac{1}{3}(\Tcal+\Scalo) & \Tcal-\Scalo \\
\end{array}
\right)\frac{Q_{0}(a)}{p}\right.\\\nonumber
&\hspace{7cm}-\left.\left(
\begin{array}{cc}
0 & -\frac{1}{3}(\Tcal+\Scalo) \\
-\frac{2}{3}(\Tcal+\Scalo) & \Tcal-\Scalo \\
\end{array}
\right)\frac{Q_1(a)}{k}\right],
\end{align}
\begin{align}
\label{eq:Wbasis3}
&{\mathbf{K_{W}}_{\mathrm{PV}}}^{\frac{1}{2}}_{1\frac{3}{2},0\frac{1}{2}}(k,p,E)=\frac{8\pi}{3}\left[\left(
\begin{array}{cc}
 \frac{2 (\Scalo-\Scalt)}{3}-2\Tcal & \frac{2\Scalo+\Scalt}{3}+2\Tcal \\
 \frac{2 (\Scalo-\Scalt)}{3}-2\Tcal & \frac{2\Scalo+\Scalt}{3}+2\Tcal \\
\end{array}
\right)\frac{Q_{0}(a)}{k}\right.\\\nonumber
&\hspace{7cm}-
\left.\left(
\begin{array}{cc}
 \frac{\Scalo-\Scalt }{3} -\Tcal& \frac{\Scalo+5\Scalt}{3}+\Tcal \\
 \frac{\Scalo-\Scalt }{3} -\Tcal & \frac{\Scalo+5\Scalt}{3}+\Tcal  \\
\end{array}
\right)\frac{Q_{1}(a)}{p}\right],
\end{align}
and
\begin{align}
\label{eq:Wbasis4}
&{\mathbf{K_{W}}_{\mathrm{PV}}}^{\frac{1}{2}}_{0\frac{1}{2},1\frac{3}{2}}(k,p,E)=\frac{8\pi}{3}\left[\left(
\begin{array}{cc}
 \frac{2 (\Scalo-\Scalt)}{3}-2\Tcal & \frac{2 (\Scalo-\Scalt)}{3}-2\Tcal \\
 \frac{2\Scalo+\Scalt}{3}+2\Tcal & \frac{2\Scalo+\Scalt}{3}+2\Tcal \\
\end{array}
\right)\frac{Q_{0}(a)}{p}\right.\\\nonumber
&\hspace{7cm}-
\left.\left(
\begin{array}{cc}
 \frac{\Scalo-\Scalt }{3} -\Tcal& \frac{\Scalo-\Scalt }{3} -\Tcal \\
 \frac{\Scalo+5\Scalt}{3}+\Tcal & \frac{\Scalo+5\Scalt}{3}+\Tcal  \\
\end{array}
\right)\frac{Q_{1}(a)}{k}\right],
\end{align}
where $a$ is given in Eq.~(\ref{eq:a}).

The asymptotic behavior of the LO PV scattering amplitude is determined by both the kernel and inhomogeneous term of Eq.~(\ref{eq:PVLOWig}).  While the kernel reproduces exactly the same asymptotic behavior as in the PC scattering amplitudes, the inhomogeneous term creates new asymptotic behavior that depends on the two-body PV LECs.  The asymptotic form of the Ws scattering amplitude for the $\DPoh\!\to\!\DS$ and $\QPoh\!\to\!\DS$ channels is given by 
\begin{align}
{t_{\mathrm{PV}}}^{\frac{1}{2};W\!s}_{0\frac{1}{2},1X}(k,q,E)=&C\frac{\sin\left(s_{0}\ln\left(\frac{q}{\Lambda^{*}}\right)\right)}{q}+C\frac{1}{\sqrt{3}}(\gamma_{t}+\gamma_{s})|B_{-1}|\frac{\sin\left(s_{0}\ln\left(\frac{q}{\Lambda^{*}}\right)+\mathrm{Arg}(B_{-1})\right)}{q^{2}}\\\nonumber
&+D^{{}^{\widehat{X}}\!P_{\nicefrac{1}{2}}}q^{-s_{1}}+D^{{}^{\widehat{X}}\!P_{\nicefrac{1}{2}}}_{-}q^{-s_{1}-1}\cdots,
\end{align}
where $X=\frac{1}{2}$ ($X=\frac{3}{2}$) for the $\DPoh\!\to\!\DS$ ($\QPoh\!\to\!\DS$) channel and $\hat{X}=2X+1$.  For the same channels the Was scattering amplitude is given by
\begin{equation}
{t_{\mathrm{PV}}}^{\frac{1}{2};W\!as}_{0\frac{1}{2},1X}(k,q,E)=-\frac{C}{2\sqrt{3}}(\gamma_{t}-\gamma_{s})|\tilde{B}_{-1}|\frac{\sin\left(s_{0}\ln\left(\frac{q}{\Lambda^{*}}\right)+\mathrm{Arg}(\tilde{B}_{-1})\right)}{q^{2}}+E^{{}^{\widehat{X}}\!P_{\nicefrac{1}{2}}}q^{-s_{1}}+\cdots.
\end{equation}
Details of the derivation of $D^{{}^{\widehat{X}}\!P_{\nicefrac{1}{2}}}$, $D_{-}^{{}^{\widehat{X}}\!P_{\nicefrac{1}{2}}}$, and $E^{{}^{\widehat{X}}\!P_{\nicefrac{1}{2}}}$ are given in Appendix~\ref{app:PVasymptotic} and their values are
\begin{equation}
\label{eq:Dvalue}
D^{{}^{\widehat{X}}\!P_{\nicefrac{1}{2}}}=\frac{\frac{1}{2}\left(a^{(X)}_{W\!s}I(1-s_{1})-b^{(X)}_{W\!s}\mathcal{M}[1,-s_{1}]\right)B^{{}^{\widehat{X}}\!P_{\nicefrac{1}{2}}}}{1-I(1-s_{1})},
\end{equation}
\begin{equation}
\label{eq:Evalue}
E^{{}^{\widehat{X}}\!P_{\nicefrac{1}{2}}}=\frac{\frac{1}{2}\left(a^{(X)}_{W\!as}\,I(1-s_{1})-b^{(X)}_{W\!as}\,\mathcal{M}[1,-s_{1}]\right)B^{{}^{\widehat{X}}\!P_{\nicefrac{1}{2}}}}{1+\frac{1}{2}I(1-s_{1})},
\end{equation}
and
\begin{align}
\label{eq:Dmvalue}
&D^{{}^{\widehat{X}}\!P_{\nicefrac{1}{2}}}_{-}=\left\{\frac{1}{2}\left(\frac{2}{\sqrt{3}}(\gamma_{t}+\bar{\delta}_{X\frac{3}{2}}\gamma_{s})B^{{}^{\widehat{X}}\!P_{\nicefrac{1}{2}}}+B^{{}^{\widehat{X}}\!P_{\nicefrac{1}{2}}}_{-1}\right)\left(a^{(X)}_{W\!s}I(-s_{1})-b^{(X)}_{W\!s}\mathcal{M}[1,-s_{1}-1]\right)\right.\\\nonumber
&\hspace{3cm}\left.+\frac{2}{\sqrt{3}}I(-s_{1})[(\gamma_{t}+\gamma_{s})D^{{}^{\widehat{X}}\!P_{\nicefrac{1}{2}}}+(\gamma_{t}-\gamma_{s})E^{{}^{\widehat{X}}\!P_{\nicefrac{1}{2}}}]\right\}/(1-I(-s_{1})),
\end{align}
where $\bar{\delta}_{X\frac{3}{2}}=1-\delta_{X\frac{3}{2}}$, 
\begin{align}
&a_{W\!s}^{(X)}=\left\{\begin{array}{ll}
-\frac{2}{3}\sqrt{2}(\Scalo+\Tcal) & X=\frac{1}{2}\\
\frac{8}{3}\left(\frac{\Scalo-\Scalt}{3}-\Tcal\right) & X=\frac{3}{2}
\end{array}\right. &&
b_{W\!s}^{(X)}=\left\{\begin{array}{ll}
-\frac{1}{3}\sqrt{2}(\Scalo+\Tcal) & X=\frac{1}{2}\\
\frac{4}{3}\left(\frac{\Scalo-\Scalt}{3}-\Tcal\right) & X=\frac{3}{2}
\end{array}\right.,
\end{align}
and
\begin{align}
\label{eq:aandbvalueWas}
&a_{W\!as}^{(X)}=\left\{\begin{array}{ll}
\sqrt{2}(\Tcal-\Scalo) & X=\frac{1}{2}\\
\frac{4}{3}\left(\frac{2\Scalo+\Scalt}{3}+2\Tcal\right) & X=\frac{3}{2}
\end{array}\right. &&
b_{W\!as}^{(X)}=\left\{\begin{array}{ll}
\sqrt{2}(\Tcal-\Scalo) & X=\frac{1}{2}\\
\frac{4}{3}\left(\frac{\Scalo+5\Scalt}{3}+\Tcal\right) & X=\frac{3}{2}
\end{array}\right..
\end{align}

The asymptotic form of the Was PV scattering amplitude for the $\DS\!\to\!\DPoh$ and $\DS\!\to\!\QPoh$ channels is given by
\begin{align}
{t_{\mathrm{PV}}}^{\frac{1}{2};W\!as}_{1X,0\frac{1}{2}}(k,q,E)=\left|H^{{}^{\widehat{X}}\!P_{\nicefrac{1}{2}}}\right|\sin\left(s_{0}\ln\left(\frac{q}{\Lambda^{*}}\right)+\mathrm{Arg}\left(H^{{}^{\widehat{X}}\!P_{\nicefrac{1}{2}}}\right)\right)+\cdots
\end{align}
where
\begin{equation}
\label{eq:Hvalue}
H^{{}^{\widehat{X}}\!P_{\nicefrac{1}{2}}}=\frac{\frac{1}{2}C\left(f^{(X)}_{W\!as}-g^{(X)}_{W\!as}\,\mathcal{M}[1,is_{0}+1]\right)}{1-\frac{1}{2}\mathcal{M}[1,is_{0}+1]},
\end{equation}
and
\begin{align}
\label{eq:fandgvalue}
&f^{(X)}_{W\!as}=\left\{\begin{array}{ll}
-\frac{2}{3}\sqrt{2}\left(\Tcal+\Scalo\right) & X=\frac{1}{2}\\
\frac{4}{3}\left(\frac{\Scalo-\Scalt}{3}-\Tcal\right) & X=\frac{3}{2}
\end{array}\right. &&
g^{(X)}_{W\!as}=\left\{\begin{array}{ll}
-\frac{1}{3}\sqrt{2}\left(\Tcal+\Scalo\right) & X=\frac{1}{2}\\
\frac{2}{3}\left(\frac{\Scalo-\Scalt}{3}-\Tcal\right) & X=\frac{3}{2}
\end{array}\right..
\end{align}
Details of the derivation of $H^{{}^{\widehat{X}}\!P_{\nicefrac{1}{2}}}$ are given in Appendix~\ref{app:PVasymptotic}.  The asymptotic form of the Ws PV scattering amplitude for these channels is not shown because it is already known or subleading compared to the Was PV scattering amplitude.  From Eq.~(\ref{eq:Wbasis1}) it is apparent that the Ws part of $\DS\!\to\!\DPoh$ scattering amplitude only picks out the Was symmetric piece of the $\DS$ PC scattering amplitude which is subleading in the asymptotic limit, and the Ws and Was PV scattering amplitudes in the $\DS\!\to\!\QPoh$ channel are identical.

\section{\label{sec:NLO}Next-to-Leading-Order Three-body Parity-Violation}
\subsection{Next-to-Leading-Order Parity-Violating three-body force}

The existence or non-existence of a PV three-body force can be addressed by looking at the asymptotic behavior of the $N\!d$ PV scattering amplitude.  If the PV scattering amplitude converges as the cutoff $\Lambda\to\infty$ then a PV three-body force is not required.  Grie{\ss}hammer and Schindler explicitly showed that the LO on-shell PV scattering amplitude converges and therefore a LO PV three-body force is not required.  However, at NLO they did not calculate the asymptotic scaling of the scattering amplitude but rather attempted to demonstrate that the only possible PV single derivative three-body force that can exist does not have the same divergence structure arising from the two-body PV contributions to the scattering amplitude and therefore no NLO PV three-body force should exist.

Using Fierz rearrangements Grie{\ss}hammer and Schindler showed that all possible single derivative PV three-body forces are equivalent to
\begin{align}
\mathcal{L}_{\mathrm{PV}}^{\mathrm{3NI}}=&\frac{y^{2}M_{N}}{3\Lambda^{2}}\left[H_{\mathrm{PV}}^{(\Delta I=0)}(\Lambda)\left[\hat{N}^{\dagger}\hat{t}_{j}^{\dagger}\sigma^{j}-\hat{N}^{\dagger}\hat{s}^{\dagger}_{A}\tau^{A}\right]\left[\hat{t}_{i}(\vec{\sigma}\cdot i\!\stackrel{\leftrightarrow}{\nabla})\sigma^{i}\hat{N}-\hat{s}_{B}(\vec{\sigma}\cdot i\!\stackrel{\leftrightarrow}{\nabla})\tau^{B}\hat{N}\right]\right.\\\nonumber
&+\left.H_{\mathrm{PV}}^{(\Delta I=1)}(\Lambda)\left[\hat{N}^{\dagger}\hat{t}_{j}^{\dagger}\sigma^{j}-\hat{N}^{\dagger}\hat{s}^{\dagger}_{A}\tau^{A}\right]\tau^{3}\left[\hat{t}_{i}(\vec{\sigma}\cdot i\!\stackrel{\leftrightarrow}{\nabla})\sigma^{i}\hat{N}-\hat{s}_{B}(\vec{\sigma}\cdot i\!\stackrel{\leftrightarrow}{\nabla})\tau^{B}\hat{N}\right]\right]+\mathrm{H.c.}
\end{align}
Projecting the tree-level PV three-body force onto the $\DS\to\DPoh$ channel gives the contribution to the scattering amplitude
\begin{equation}
i\mathbf{K}^{\frac{1}{2};\mathrm{3NI}}_{1\frac{1}{2},0\frac{1}{2}}(k,p,E)=A_{\mathrm{3NI}}\left(H_{\mathrm{PV}}^{(\Delta I=0)}(\Lambda)+\tau^{3}H_{\mathrm{PV}}^{(\Delta I=1)}(\Lambda)\right)\left(\!\!\begin{array}{rr}
1 & -1 \\[-2mm]
-1 & 1
\end{array}\!\!\right),
\end{equation}
where $A_{\mathrm{3NI}}$ contains all the momentum dependence which is not essential in the proceeding.  Projecting the c.c.~space matrix into the Wigner basis gives
\begin{align}
i\mathbf{K_{W}}^{\frac{1}{2};\mathrm{3NI}}_{1\frac{1}{2},0\frac{1}{2}}(k,p,E)&=\frac{1}{2}\left(\!\!\begin{array}{rr}
1 & -1 \\[-2mm]
1 & 1
\end{array}\!\!\right)i\mathbf{K}^{\frac{1}{2};\mathrm{3NI}}_{1\frac{1}{2},0\frac{1}{2}}(k,p,E)\left(\!\!\begin{array}{rr}
1 & 1 \\[-2mm]
-1 & 1
\end{array}\!\!\right)\\\nonumber
&=A_{\mathrm{3NI}}\left(H_{\mathrm{PV}}^{(\Delta I=0)}(\Lambda)+\tau^{3}H_{\mathrm{PV}}^{(\Delta I=1)}(\Lambda)\right)\left(\!\!\begin{array}{cc}
2 & 0 \\[-2mm]
0 & 0
\end{array}\!\!\right).
\end{align}
Thus, the only possible single derivative PV three-body force connects only Ws amplitudes to Ws amplitudes.  The PC scattering amplitudes that are integrated with the tree level PV three-body force diagram to get the full contribution to the scattering amplitude from the PV three-body force are diagonal in the Wigner-basis.\footnote{The NLO PC scattering amplitudes are diagonal in the Wigner-basis only if $c_{0t}^{(0)}=c_{0s}^{(0)}$, which is nearly the case.  The difference $\frac{1}{2}(c_{0t}^{(0)}-c_{0s}^{(0)})$ can be treated as a perturbative correction to the value $\frac{1}{2}(c_{0t}^{(0)}+c_{0s}^{(0)})$~\cite{Griesshammer:2010nd}.}   Therefore, the full scattering amplitude due to the PV three-body force only has a Ws to Ws piece in the Wigner basis.

Meanwhile, projecting out the PV two-nucleon tree level diagrams into the $\DS\to\DPoh$ channel and Wigner basis gives Eq.~(\ref{eq:Wbasis1}).  Since the PC amplitudes which must be convoluted with the two-nucleon PV tree level diagrams are diagonal in the Wigner basis, and the tree-level diagrams do not contain a Ws to Ws term, as seen in Eq.~(\ref{eq:Wbasis1}), the full scattering amplitude due to two-nucleon PV interactions will not contain a Ws to Ws contribution.  However, this is the only part of c.c.~space that the PV three-body force occupies and therefore it cannot renormalize divergences from the NLO PV two-nucleon diagrams.  This is the argument made by Grie{\ss}hammer and Schindler~\cite{Griesshammer:2010nd}.

By Fierz rearrangement, it can be shown that~\cite{Griesshammer:2010nd}
\begin{equation}
\sigma^{i}\hat{t}_{i}\hat{N}^{b}=-\left(\tau^{A}\right)^{b}_{\,\,\,\,c}\hat{s}_{A}\hat{N}^{c},
\end{equation}
and
\begin{equation}
\hat{t}_{i}(\vec{\sigma}\cdot\stackrel{\leftrightarrow}{\nabla})\sigma^{i}\hat{N}^{b}=-\hat{s}_{A}(\vec{\sigma}\cdot\stackrel{\leftrightarrow}{\nabla})(\tau_{A})^{b}_{\,\,\,\,c}\hat{N}^{c}.
\end{equation}
This allows the form of the single derivative three-nucleon force to be written as
\begin{align}
\mathcal{L}_{\mathrm{PV}}^{\mathrm{3NI}}=&\frac{y^{2}M_{N}}{3\Lambda^{2}}\left[\hat{N}^{\dagger}\hat{t}_{j}^{\dagger}\sigma^{j}\right]\left(H_{\mathrm{PV}}^{(\Delta I=0)}(\Lambda)+\tau^{3}H_{\mathrm{PV}}^{(\Delta I=1)}(\Lambda)\right)\left[\hat{t}_{i}(\vec{\sigma}\cdot i\!\stackrel{\leftrightarrow}{\nabla})\sigma^{i}\hat{N}\right],
\end{align}
which projected onto the $\DS\to\DPoh$ channel and Wigner basis gives
\begin{align}
i\mathbf{K_{W}}^{\frac{1}{2};\mathrm{3NI}}_{1\frac{1}{2},0\frac{1}{2}}(k,p,E)&=\frac{1}{2}\left(\!\!\begin{array}{rr}
1 & -1 \\[-2mm]
1 & 1
\end{array}\!\!\right)i\mathbf{K}^{\frac{1}{2};\mathrm{3NI}}_{1\frac{1}{2},0\frac{1}{2}}(k,p,E)\left(\!\!\begin{array}{rr}
1 & 1 \\[-2mm]
-1 & 1
\end{array}\!\!\right)\\\nonumber
&=\frac{1}{2}A_{\mathrm{3NI}}\left(H_{\mathrm{PV}}^{(\Delta I=0)}(\Lambda)+\tau^{3}H_{\mathrm{PV}}^{(\Delta I=1)}(\Lambda)\right)\left(\!\!\begin{array}{rr}
1 & 1 \\[-2mm]
1 & 1
\end{array}\!\!\right).
\end{align}
It is immediately obvious that this form of the PV three-body force has a structure in c.c.~space that will overlap with the structure of the scattering amplitudes due to the two-nucleon PV contributions to PV $N\!d$ scattering.  Thus, the argument made by Grie{\ss}hammer and Schindler does not hold since the Wigner-basis structure of an operator is not invariant under Fierz rearrangement.  This does not mean that a PV three-body force must exist, but only that it could.  To show the existence or non-existence of such a NLO PV three-body force a proper asymptotic analysis of the NLO PV scattering amplitude must be carried out, which is done below.

\subsection{Next-to-Leading-Order Asymptotics}

\begin{figure}[hbt]
\includegraphics[width=120mm]{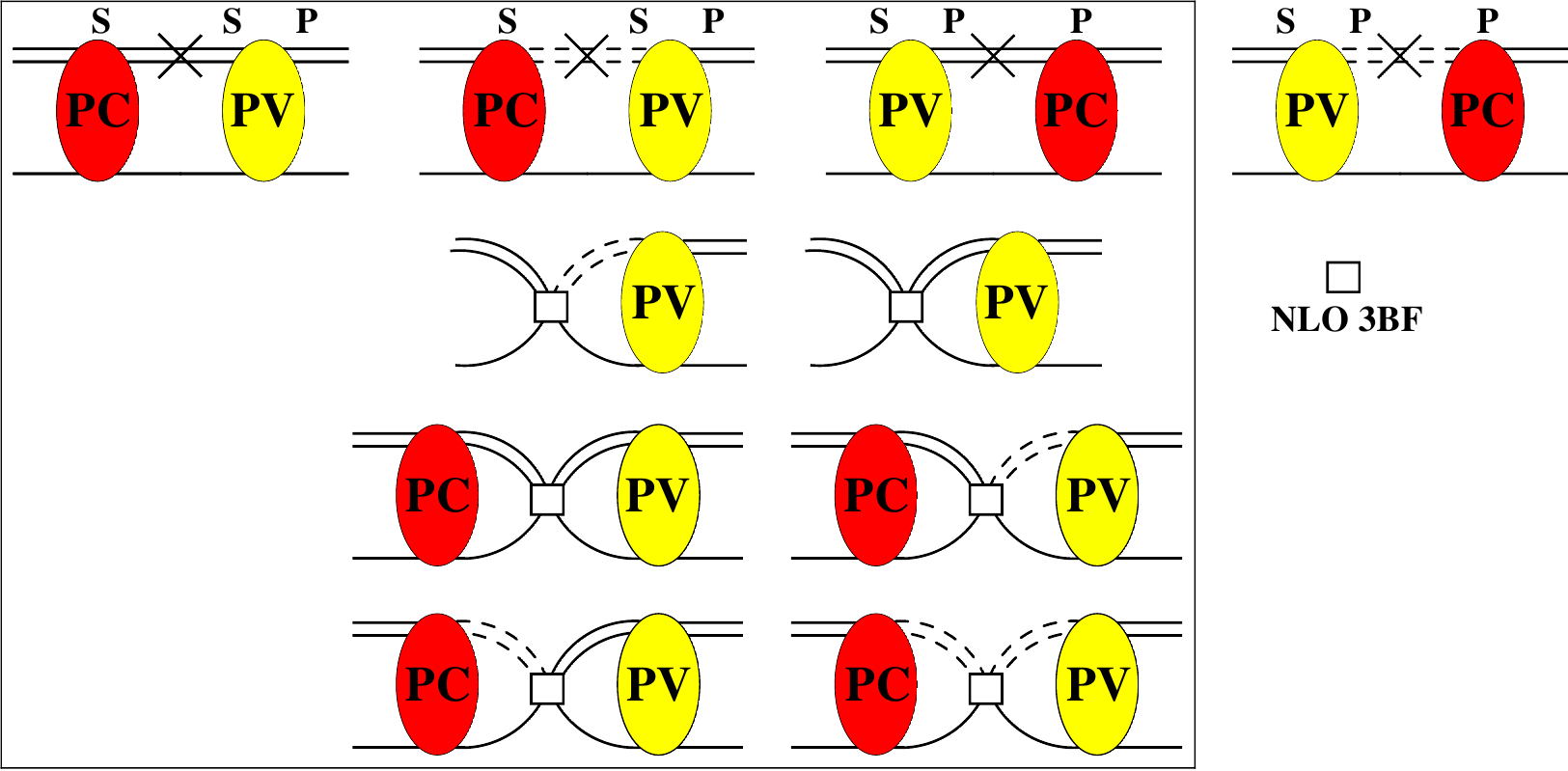}
\caption{\label{fig:PVNLODiagrams} Diagrams for the NLO PV scattering amplitude.  The cross represents an effective range insertion and the empty box an instance of the NLO PC three-body force.  Diagrams in the enclosed area occur for both the $\DS\!\to\!\DPoh$ and $\DS\!\to\!\QPoh$ channels, but the diagram outside of the enclosed area only occurs in the $\DS\!\to\!\DPoh$ channel.}
\end{figure}

The NLO PV $N\!d$ scattering amplitude in the $\DS\!\to\DPoh$ and $\DS\to\QPoh$ channels is given by the set of diagrams in Fig.~\ref{fig:PVNLODiagrams}, 
where the cross represents an effective range insertion and the empty box an insertion of the NLO PC three-body force fit to the doublet $S$-wave $nd$ scattering length~\cite{Vanasse:2015fph}.\footnote{The NLO $N\!d$ PV scattering amplitude can also be calculated using integral equations following the methods of Ref.~\cite{Vanasse:2013sda}, which gives results in agreement with the summation of diagrams in Fig.~\ref{fig:PVNLODiagrams}.}  Diagrams inside the enclosed area occur for both channels, while the diagram outside the enclosed area does not occur in the $\DS\!\to\!\QPoh$ channel.  Summing these diagrams gives 
\begin{align}
\label{eq:NLOPVamp}
{t^{\mathrm{NLO}}_{\mathrm{PV}}}^{J;Nt\to Nt}_{L'S',LS}&(k,k,E)=\\\nonumber
&\left[{\mathbf{t}_{\mathrm{PV}}}^{J}_{LS,L'S'}(k,q,E)\right]^{T}\otimes\mathbf{D}_{\mathrm{NLO}}\left(E,q\right){\mathbf{t}_{\mathrm{PC}}}^{J}_{LS,LS}(k,q,E)\\\nonumber
&+\left[{\mathbf{t}_{\mathrm{PC}}}^{J}_{L'S',L'S'}(k,q,E)\right]^{T}\otimes\mathbf{D}_{\mathrm{NLO}}\left(E,q\right){\mathbf{t}_{\mathrm{PV}}}^{J}_{L'S',LS}(k,q,E)\\\nonumber
&+\frac{4\pi H_{\mathrm{NLO}}(\Lambda)}{\Lambda^{2}}\left[{\mathbf{t}_{\mathrm{PV}}}^{J}_{LS,L'S'}(k,q,E)\right]^{T}\otimes\mathbf{D}\left(E,q\right)\\\nonumber
&\times\left(\!\!\begin{array}{rr}
1 & -1\\[-2mm]
-1 & 1
\end{array}\right)\left(\mathbf{v}_{p}+\mathbf{D}\left(E,\ell\right)\otimes{\mathbf{t}_{\mathrm{PC}}}^{J}_{LS,LS}(k,\ell,E)\right),
\end{align}
for the on-shell NLO PV scattering amplitude, where the NLO dibaryon propagators are defined by the c.c.~space matrix
\begin{equation}
\mathbf{D}_{\mathrm{NLO}}(E,q)=\left(\begin{array}{cc}
\frac{Z_{t}-1}{2\gamma_{t}}\frac{\sqrt{\frac{3}{4}q^{2}-M_{N}E-i\epsilon}+\gamma_{t}}{\sqrt{\frac{3}{4}q^{2}-M_{N}E-i\epsilon}-\gamma_{t}} & 0 \\
0 & \frac{Z_{s}-1}{2\gamma_{s}}\frac{\sqrt{\frac{3}{4}q^{2}-M_{N}E-i\epsilon}+\gamma_{s}}{\sqrt{\frac{3}{4}q^{2}-M_{N}E-i\epsilon}-\gamma_{s}}\end{array}\right).
\end{equation}
To get the asymptotic form of the NLO PV scattering amplitude it is transformed to the Wigner basis and the asymptotic forms of the dibaryon propagators, $H_{\mathrm{NLO}}(\Lambda)$, and the LO PC and PV scattering amplitudes are plugged in keeping only those pieces that are divergent in the limit $\Lambda\to\infty$.  Doing this for the $\DS\!\to\!\DPoh$ channel gives
\begin{align}
\label{eq:NLOasymptoticDSDP}
{t^{\mathrm{NLO}}_{\mathrm{PV}}}^{\frac{1}{2};Nt\to Nt}_{1\frac{1}{2},0\frac{1}{2}}(k,k,E)_{\mathrm{div}}=&\frac{1}{16\pi^{2}}\left(\rho_{t}+\rho_{s}\right)\int_{\Lambda}dqq^{2}{t^{(-1)}_{\mathrm{PC}}}^{\frac{1}{2};W\!s}_{0\frac{1}{2},0\frac{1}{2}}(k,q,E){t_{\mathrm{PV}}^{(-1.7)}}^{\frac{1}{2};W\!s}_{0\frac{1}{2},1\frac{1}{2}}(k,q,E)\\\nonumber
&+\frac{1}{16\pi^{2}}\left(\rho_{t}-\rho_{s}\right)\int_{\Lambda}dqq^{2}{t_{\mathrm{PC}}^{(-1)}}^{\frac{1}{2};W\!s}_{0\frac{1}{2},0\frac{1}{2}}(k,q,E){t_{\mathrm{PV}}^{(-1.7)}}^{\frac{1}{2};W\!as}_{0\frac{1}{2},1\frac{1}{2}}(k,q,E)\\\nonumber
&+\frac{1}{16\pi^{2}}\left(\rho_{t}+\rho_{s}\right)\int_{\Lambda}dqq^{2}{t_{\mathrm{PV}}^{(0)}}^{\frac{1}{2};W\!as}_{1\frac{1}{2},0\frac{1}{2}}(k,q,E){t_{\mathrm{PC}}^{(-2.7)}}^{\frac{1}{2};W\!as}_{1\frac{1}{2},1\frac{1}{2}}(k,q,E)\\\nonumber
&+\frac{4H_{\mathrm{NLO}}(\Lambda)}{3\pi^{3}\Lambda^{2}}\int_{\Lambda}dqq{t_{\mathrm{PC}}^{(-1)}}^{\frac{1}{2};W\!s}_{0\frac{1}{2},0\frac{1}{2}}(k,q,E)\int_{\Lambda}d\ell\ell{t_{\mathrm{PV}}^{(-1.7)}}^{\frac{1}{2};W\!s}_{0\frac{1}{2},1\frac{1}{2}}(k,\ell,E),
\end{align}
where the superscript in parentheses refers to the part of the scattering amplitude with that power law scaling in the asymptotic limit.  For convenience
\begin{equation}
\frac{1}{2}\rho_{t}=\frac{Z_{t}-1}{2\gamma_{t}}\quad,\quad\frac{1}{2}\rho_{s}=\frac{Z_{s}-1}{2\gamma_{s}},
\end{equation}
is defined.  Plugging in the asymptotic forms for the scattering amplitudes into Eq.~(\ref{eq:NLOasymptoticDSDP}) and solving the resulting indefinite integrals gives
\begin{align}
&{t^{\mathrm{NLO}}_{\mathrm{PV}}}^{\frac{1}{2};Nt\to Nt}_{1\frac{1}{2},0\frac{1}{2}}(k,k,E)_{\mathrm{div}}=\\\nonumber
&\frac{1}{16\pi^{2}}\frac{\Lambda^{2-s_{1}}}{\sqrt{s_{0}^{2}+(s_{1}-2)^{2}}}\left[(\rho_{t}+\rho_{s})\left\{CD^{\DPoh}\sin\left(s_{0}\ln\left(\frac{\Lambda}{\Lambda^{*}}\right)+\arctan\left(\frac{s_{0}}{s_{1}-2}\right)\right)\right.\right.\\\nonumber
&\left.\left.\hspace{2cm}+B^{\DPoh}|H^{\DPoh}|\sin\left(s_{0}\ln\left(\frac{\Lambda}{\Lambda^{*}}\right)+\arctan\left(\frac{s_{0}}{s_{1}-2}\right)+\mathrm{Arg}\left(H^{\DPoh}\right)\right)\right\}\right.\\\nonumber
&\left.\hspace{2cm}+(\rho_{t}-\rho_{s})CE^{\DPoh}\sin\left(s_{0}\ln\left(\frac{\Lambda}{\Lambda^{*}}\right)+\arctan\left(\frac{s_{0}}{s_{1}-2}\right)\right)\right]\\\nonumber
&\hspace{2cm}+\frac{4H_{\mathrm{NLO}}(\Lambda)}{3\pi^{3}\Lambda^{2}}CD^{\DPoh}\frac{1}{\sqrt{1+s_{0}^{2}}(2-s_{1})}\Lambda^{3-s_{1}}\sin\left(s_{0}\ln\left(\frac{\Lambda}{\Lambda^{*}}\right)-\arctan(s_{0})\right)+b,
\end{align}
for the NLO scattering amplitude, where the leading contribution to $H_{\mathrm{NLO}}(\Lambda)$, calculated previously~\cite{Hammer:2001gh,Ji:2011qg,Vanasse:2014kxa}, is
\begin{equation}
H_{\mathrm{NLO}}(\Lambda)=-\Lambda\frac{3\pi(1+s_{0}^{2})}{128}(\rho_{t}+\rho_{s})\frac{\left(1-\frac{1}{\sqrt{1+4s_{0}^{2}}}\sin\left(2s_{0}\ln\left(\frac{\Lambda}{\Lambda^{*}}\right)+\arctan\left(\frac{1}{2s_{0}}\right)\right)\right)}{\sin^{2}\left(s_{0}\ln\left(\frac{\Lambda}{\Lambda^{*}}\right)-\arctan(s_{0})\right)}+\cdots.
\end{equation}
The resulting indefinite integrals are only evaluated at their upper limit $\Lambda$ and not at zero momentum.  At zero momentum the form of the scattering amplitudes is not known analytically, and this unknown IR physics is encapsulated in the constant $b$ which is obtained by fitting the asymptotic form to the numerically calculated scattering amplitude.

Going to the Wigner basis and keeping divergent pieces for the $\DS\to\QPoh$ channel gives
\begin{align}
\label{eq:NLOasymptoticDSQP}
{t^{\mathrm{NLO}}_{\mathrm{PV}}}^{\frac{1}{2};Nt\to Nt}_{1\frac{3}{2},0\frac{1}{2}}&(k,k,E)_{\mathrm{div}}=\frac{1}{16\pi^{2}}\left(\rho_{t}+\rho_{s}\right)\int_{\Lambda}dqq^{2}{t^{(-1)}_{\mathrm{PC}}}^{\frac{1}{2};W\!s}_{0\frac{1}{2},0\frac{1}{2}}(k,q,E){t_{\mathrm{PV}}^{(-1.7)}}^{\frac{1}{2};W\!s}_{0\frac{1}{2},1\frac{3}{2}}(k,q,E)\\\nonumber
&+\frac{1}{16\pi^{2}}\left(\rho_{t}-\rho_{s}\right)\int_{\Lambda}dqq^{2}{t_{\mathrm{PC}}^{(-1)}}^{\frac{1}{2};W\!s}_{0\frac{1}{2},0\frac{1}{2}}(k,q,E){t_{\mathrm{PV}}^{(-1.7)}}^{\frac{1}{2};W\!as}_{0\frac{1}{2},1\frac{3}{2}}(k,q,E)\\\nonumber
&+\frac{1}{16\pi^{2}}\rho_{t}\int_{\Lambda}dqq^{2}{t_{\mathrm{PV}}^{(0)}}^{\frac{1}{2};W\!as}_{1\frac{1}{2},0\frac{1}{2}}(k,q,E){t_{\mathrm{PC}}^{(-2.7)}}^{\frac{1}{2};W\!as}_{1\frac{3}{2},1\frac{3}{2}}(k,q,E)\\\nonumber
&+\frac{4H_{\mathrm{NLO}}(\Lambda)}{3\pi^{3}\Lambda^{2}}\int_{\Lambda}dqq{t_{\mathrm{PC}}^{(-1)}}^{\frac{1}{2};W\!s}_{0\frac{1}{2},0\frac{1}{2}}(k,q,E)\int_{\Lambda}d\ell\ell{t_{\mathrm{PV}}^{(-1.7)}}^{\frac{1}{2};W\!s}_{0\frac{1}{2},1\frac{3}{2}}(k,\ell,E)\\\nonumber
&+\frac{4H_{\mathrm{NLO}}(\Lambda)}{3\pi^{3}\Lambda^{2}}\int_{\Lambda}dqq{t_{\mathrm{PC}}^{(-1)}}^{\frac{1}{2};W\!s}_{0\frac{1}{2},0\frac{1}{2}}(k,q,E)\int_{\Lambda}d\ell\ell{t_{\mathrm{PV}}^{(-2.7)}}^{\frac{1}{2};W\!s}_{0\frac{1}{2},1\frac{3}{2}}(k,\ell,E)\\\nonumber
&+\frac{2}{\sqrt{3}}(\gamma_{t}-\gamma_{s})\frac{4H_{\mathrm{NLO}}(\Lambda)}{3\pi^{3}\Lambda^{2}}\int_{\Lambda}dqq{t_{\mathrm{PC}}^{(-1)}}^{\frac{1}{2};W\!s}_{0\frac{1}{2},0\frac{1}{2}}(k,q,E)\int_{\Lambda}d\ell{t_{\mathrm{PV}}^{(-1.7)}}^{\frac{1}{2};W\!as}_{0\frac{1}{2},1\frac{3}{2}}(k,\ell,E)\\\nonumber
&+\frac{2}{\sqrt{3}}(\gamma_{t}+\gamma_{s})\frac{4H_{\mathrm{NLO}}(\Lambda)}{3\pi^{3}\Lambda^{2}}\int_{\Lambda}dqq{t_{\mathrm{PC}}^{(-1)}}^{\frac{1}{2};W\!s}_{0\frac{1}{2},0\frac{1}{2}}(k,q,E)\int_{\Lambda}d\ell{t_{\mathrm{PV}}^{(-1.7)}}^{\frac{1}{2};W\!s}_{0\frac{1}{2},1\frac{3}{2}}(k,\ell,E).
\end{align}
The last three pieces are not strictly divergent in the $\Lambda\to\infty$ limit, however, they give rise to a series of log periodic first order poles that will be noticeable at sizable, but smaller cutoffs.  Plugging in the asymptotic forms for the scattering amplitudes and solving the resulting indefinite integrals as before gives
\begin{align}
\label{eq:PVNLOaympQP}
&{t^{\mathrm{NLO}}_{\mathrm{PV}}}^{\frac{1}{2};Nt\to Nt}_{1\frac{3}{2},0\frac{1}{2}}(k,k,E)_{\mathrm{div}}=\\\nonumber
&\hspace{1cm}\frac{1}{16\pi^{2}}\frac{\Lambda^{2-s_{1}}}{\sqrt{s_{0}^{2}+(s_{1}-2)^{2}}}\left[(\rho_{t}+\rho_{s})CD^{\QPoh}\sin\left(s_{0}\ln\left(\frac{\Lambda}{\Lambda^{*}}\right)+\arctan\left(\frac{s_{0}}{s_{1}-2}\right)\right)\right.\\\nonumber
&\left.\hspace{2cm}+(\rho_{t}-\rho_{s})CE^{\QPoh}\sin\left(s_{0}\ln\left(\frac{\Lambda}{\Lambda^{*}}\right)+\arctan\left(\frac{s_{0}}{s_{1}-2}\right)\right)\right.\\\nonumber
&\left.\hspace{2cm}+4\rho_{t}B^{\QPoh}|H^{\QPoh}|\sin\left(s_{0}\ln\left(\frac{\Lambda}{\Lambda^{*}}\right)+\arctan\left(\frac{s_{0}}{s_{1}-2}\right)+\mathrm{Arg}\left(H^{\QPoh}\right)\right)\right]\\\nonumber
&\hspace{2cm}+\frac{4H_{\mathrm{NLO}}(\Lambda)}{3\pi^{3}\Lambda^{2}}CD^{\QPoh}\frac{1}{\sqrt{1+s_{0}^{2}}(2-s_{1})}\Lambda^{3-s_{1}}\sin\left(s_{0}\ln\left(\frac{\Lambda}{\Lambda^{*}}\right)-\arctan(s_{0})\right)\\\nonumber
&\hspace{2cm}+\frac{4H_{\mathrm{NLO}}(\Lambda)}{3\pi^{3}\Lambda^{2}}CD^{\QPoh}\frac{\frac{2}{\sqrt{3}}(\gamma_{t}+\gamma_{s})}{\sqrt{1+s_{0}^{2}}(1-s_{1})}\Lambda^{2-s_{1}}\sin\left(s_{0}\ln\left(\frac{\Lambda}{\Lambda^{*}}\right)-\arctan(s_{0})\right)\\\nonumber
&\hspace{2cm}+\frac{4H_{\mathrm{NLO}}(\Lambda)}{3\pi^{3}\Lambda^{2}}CD^{\QPoh}_{-}\frac{1}{\sqrt{1+s_{0}^{2}}(1-s_{1})}\Lambda^{2-s_{1}}\sin\left(s_{0}\ln\left(\frac{\Lambda}{\Lambda^{*}}\right)-\arctan(s_{0})\right)\\\nonumber
&\hspace{2cm}+\frac{4H_{\mathrm{NLO}}(\Lambda)}{3\pi^{3}\Lambda^{2}}CE^{\QPoh}\frac{\frac{2}{\sqrt{3}}(\gamma_{t}-\gamma_{s})}{\sqrt{1+s_{0}^{2}}(1-s_{1})}\Lambda^{2-s_{1}}\sin\left(s_{0}\ln\left(\frac{\Lambda}{\Lambda^{*}}\right)-\arctan(s_{0})\right)+b,
\end{align}
for the NLO scattering amplitude in the $\DS\!\to\QPoh$ channel.  The constant $b$ is fit to the cutoff dependence of the on-shell scattering amplitude for $k=1$~MeV and the resulting values for different channels and two-body PV LECs are given in Table~\ref{tab:btable} .
\begin{table}[hbt]
\begin{tabular}{|c|c|c|c|}
\hline
LEC & LEC & $\DS\!\to\DPoh$ & $\DS\!\to\!\QPoh$ \\\hline
$g_{1}$ & $\frac{\Scalo+2\Scalt}{9}$ & 0.00267~MeV$^{-2}$ & -0.001078~MeV$^{-2}$\\
$g_{2}$ & $\frac{\Scalo-\Scalt}{3\tau_{3}}$ &-0.00178~MeV$^{-2}$ & \phantom{-}0.0019\phantom{00}~MeV$^{-2}$ \\
$g_{3}+\frac{2}{3}\tau_{3}g_{4}$ & $\frac{1}{3}\Tcal$ & -0.00160~MeV$^{-2}$& -0.0019\phantom{00}~MeV$^{-2}$\\\hline
\end{tabular}
\caption{\label{tab:btable}Value of $b$ fit to the on-shell NLO PV scattering amplitude with $k=1$~MeV for $\DS\!\to\!\DPoh$ and $\DS\!\to\!\QPoh$ channels for respective combinations of the two-body PV LECs.  The factor of $\tau_{3}$ is +1 (-1) for $pd$ ($nd$) scattering.}
\end{table}

\begin{figure}[H]
	\begin{center}
	\begin{tabular}{cc}
	\includegraphics[width=90mm]{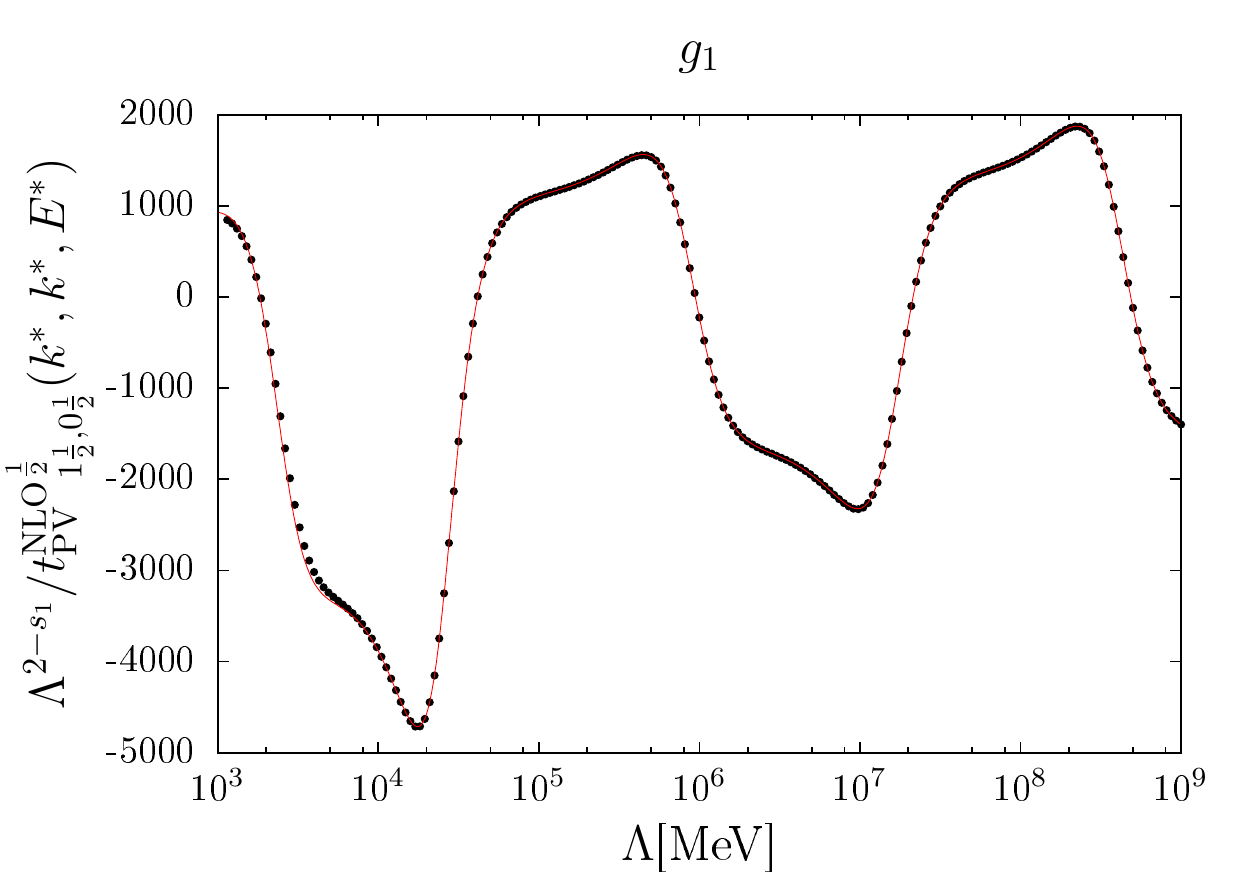} &
	\includegraphics[width=90mm]{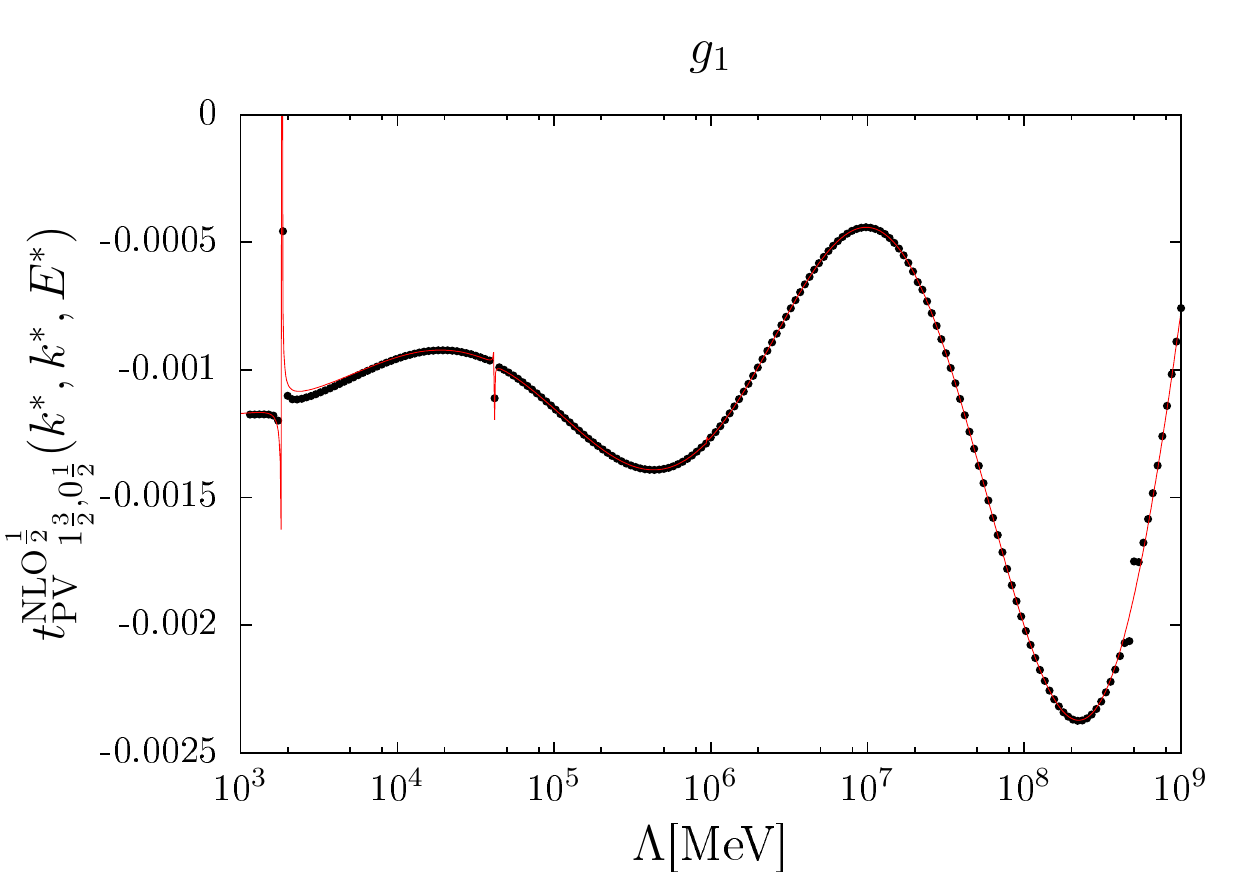} \\
	\includegraphics[width=90mm]{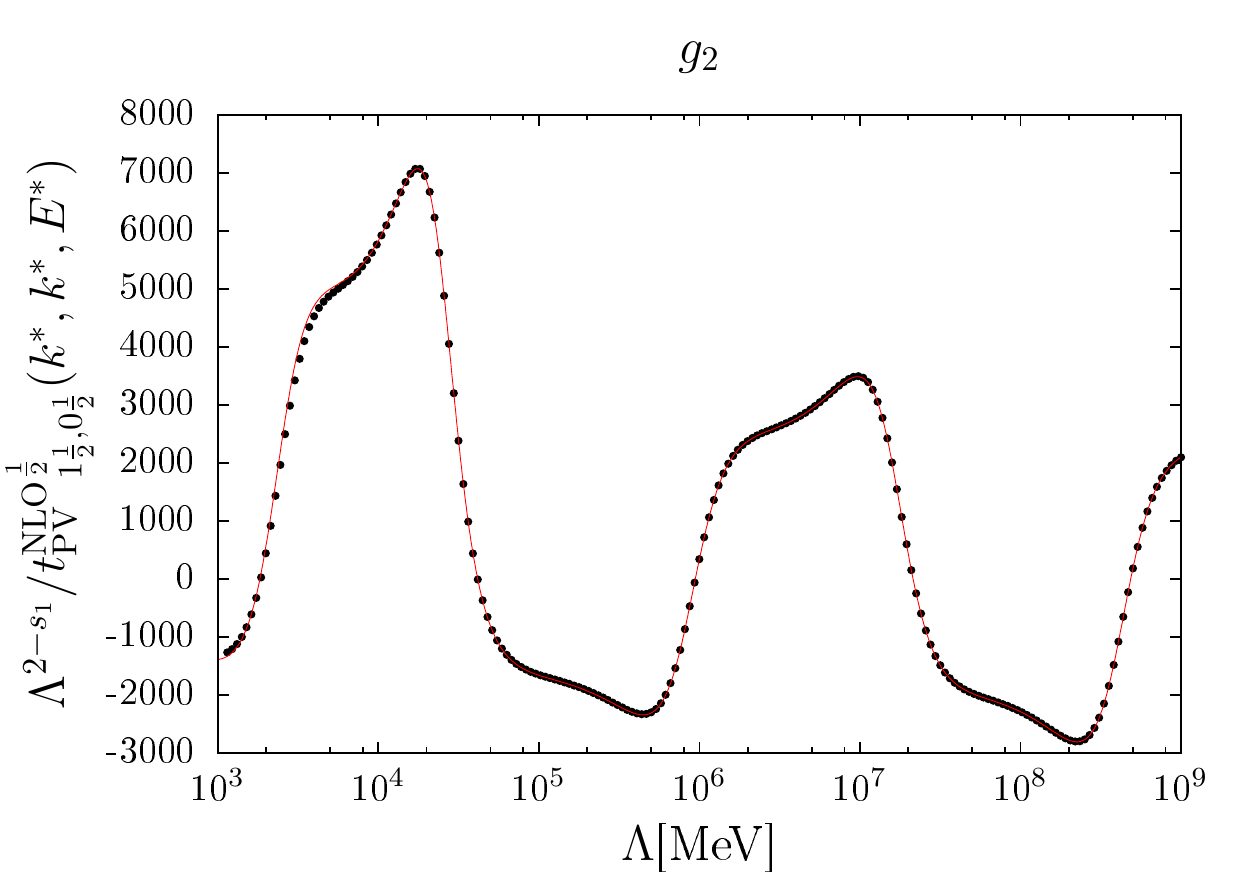} &
	\includegraphics[width=90mm]{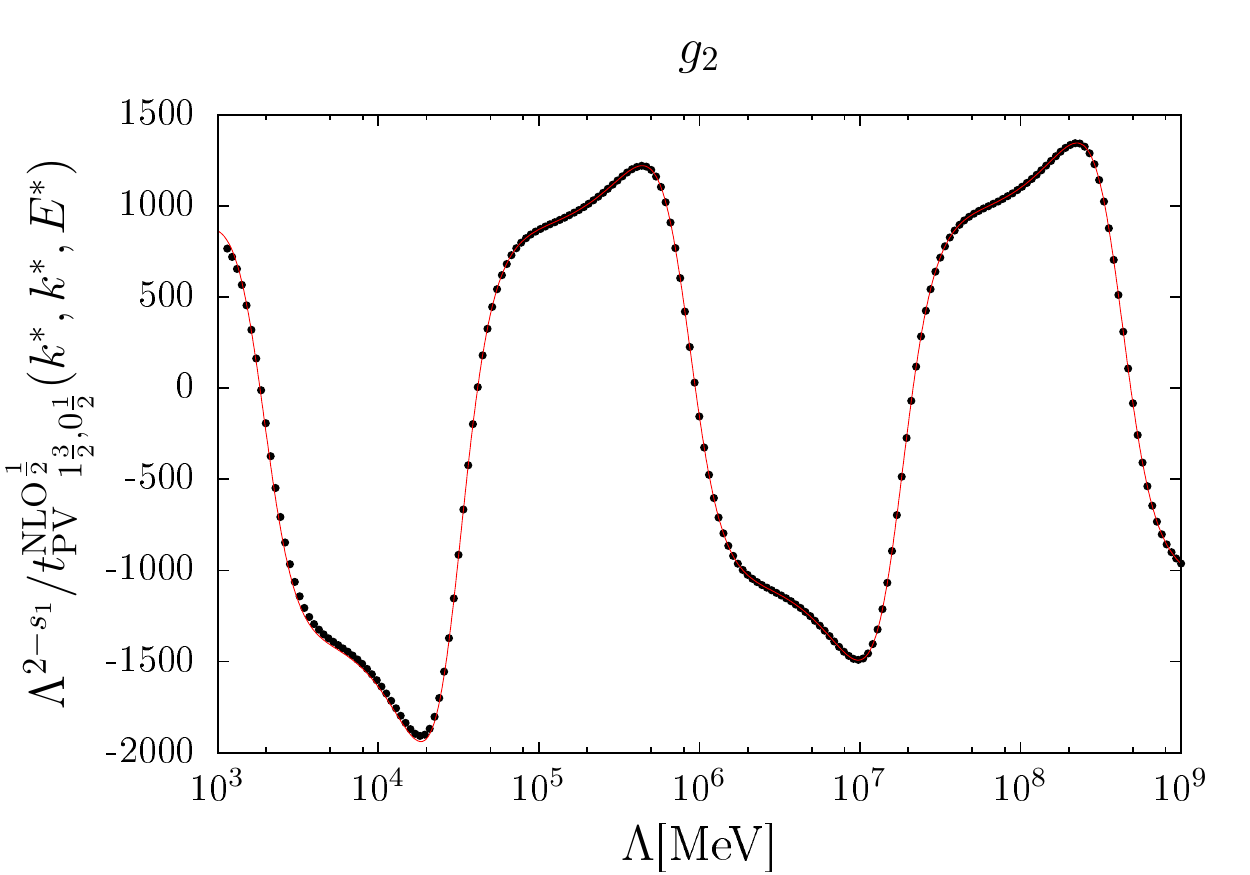} \\
	\includegraphics[width=90mm]{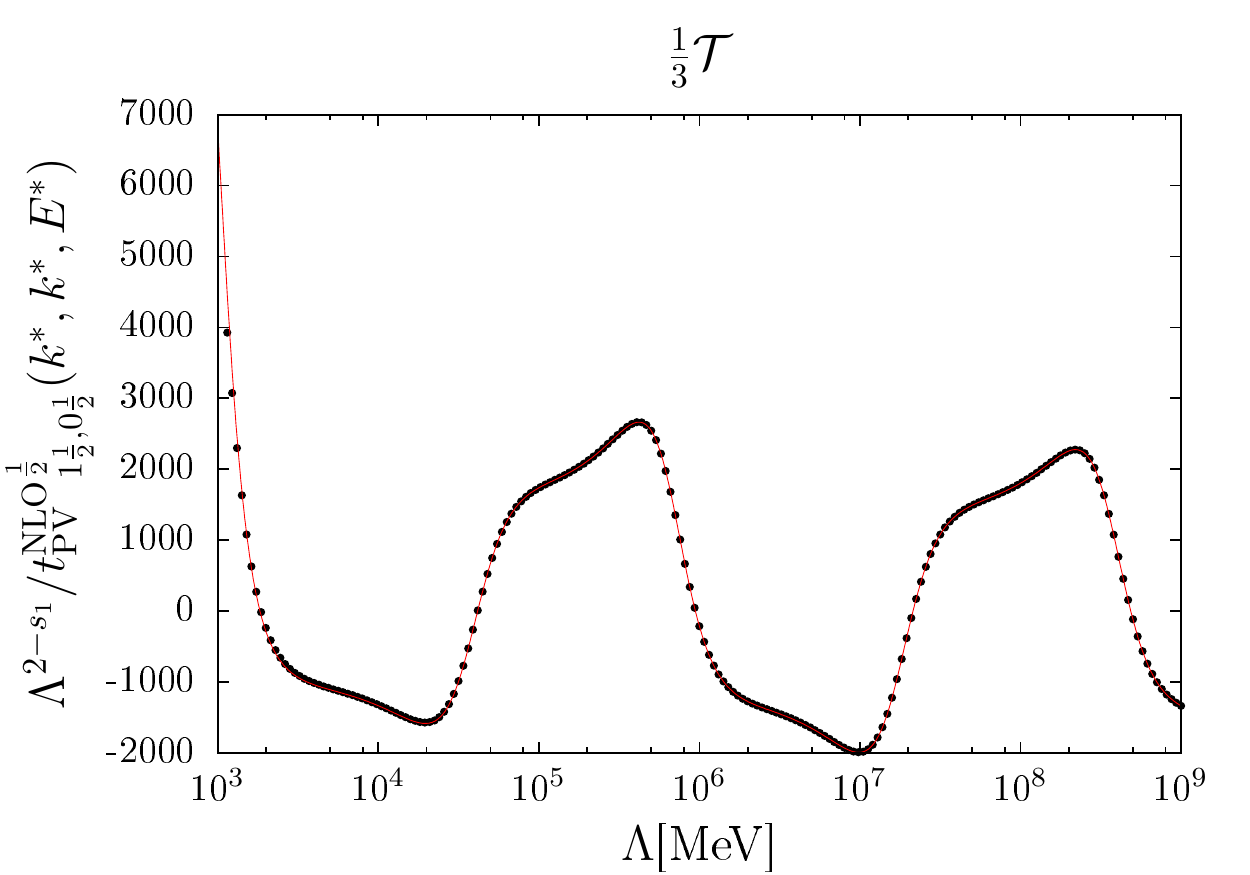} &
	\includegraphics[width=90mm]{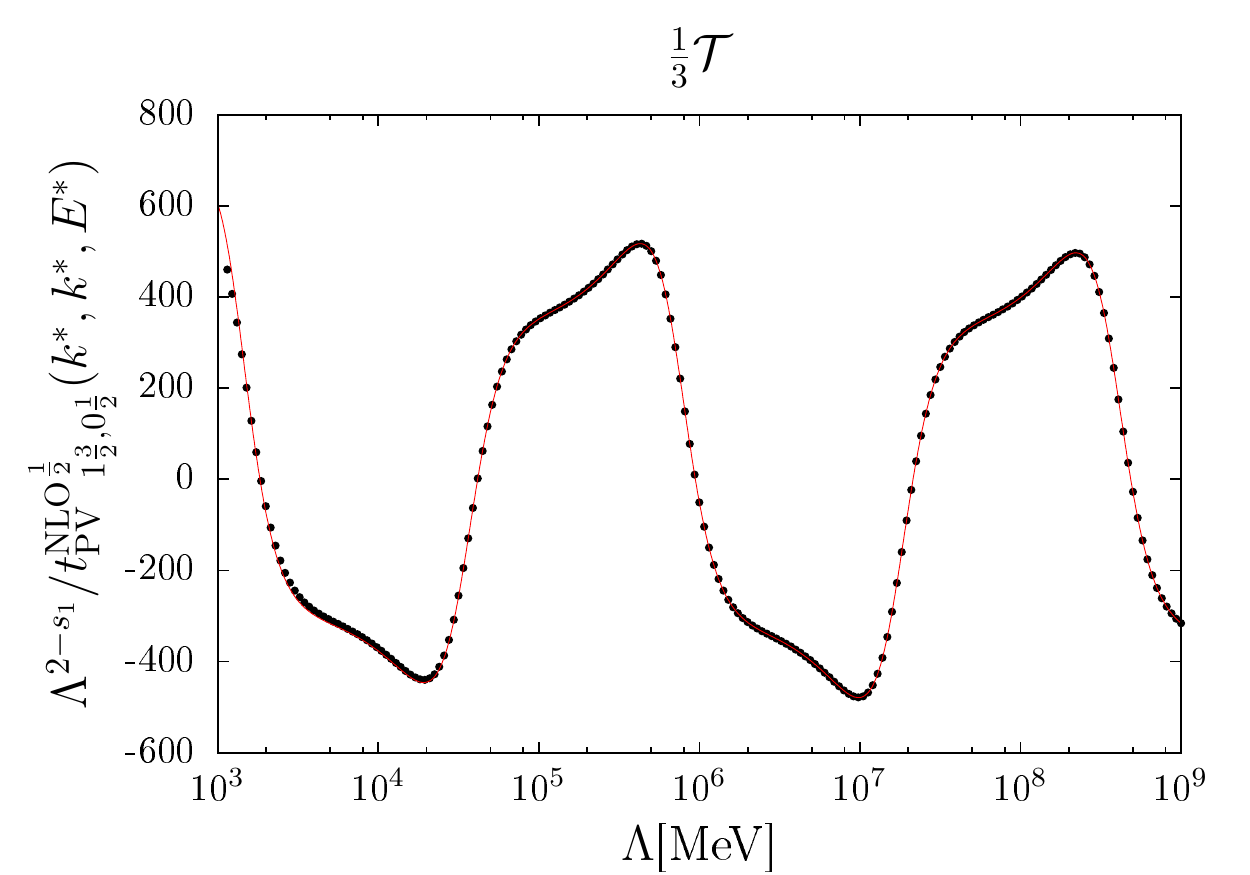} \\
	\end{tabular}

	\end{center}

\caption{\label{fig:NLOAsymptotics}Asymptotic behavior of the NLO PV scattering amplitudes for two-body PV LECs.  The black dots represent the numerical calculations and the red lines the analytical predictions.  Plots on the left (right) are for the $\DS\!\to\!\DPoh$ ($\DS\!\to\!\QPoh$) channel.  For each plot, the titled LEC is set to one and all other LECs to zero. $k^{*}=1$~MeV and $E^{*}=\frac{3}{4}\frac{(k^{*})^{2}}{M_{N}}-\frac{\gamma_{t}^{2}}{M_{N}}$. }
\end{figure}
The NLO PV scattering amplitude is calculated numerically using Eq.~(\ref{eq:NLOPVamp}).  To avoid finite $\Lambda$ effects the LO PC and PV amplitudes are calculated at a large cutoff $\Lambda=10^{12}$~MeV and then using these the integrals in Eq.~(\ref{eq:NLOPVamp}), and $H_{\mathrm{NLO}}(\Lambda)$ using the methods in Ref.~\cite{Vanasse:2014kxa}, are calculated at smaller cutoffs $\bar{\Lambda}$ up to $\bar{\Lambda}=10^{9}$~MeV.  The numerically calculated cutoff dependence of the NLO PV on-shell scattering amplitudes for $k=1$~MeV is compared to the analytical asymptotic behavior in Fig.~\ref{fig:NLOAsymptotics}.  Each plot sets one of the two-body PV LECs to one and all others to zero.  To convert poles from $H_{\mathrm{NLO}}(\Lambda)$ to zeros and divide out the dominant $\Lambda^{2-s_{1}}$ behavior, $\Lambda^{2-s_{1}}$ divided by the NLO PV scattering amplitude is plotted in Fig.~\ref{fig:NLOAsymptotics}.  However, for the LEC $g_{1}$, $D^{\QPoh}$ and $H^{\QPoh}$ are zero, which makes the subleading behavior from $H_{\mathrm{NLO}}(\Lambda)$, from the last two lines of Eq.~(\ref{eq:PVNLOaympQP}), apparent for small cutoffs .  Thus for $g_{1}$ in the $\DS\!\to\QPoh$ channel the NLO PV scattering amplitude is plotted in Fig.~\ref{fig:NLOAsymptotics}, and subleading behavior from $H_{\mathrm{NLO}}(\Lambda)$ leads to first order poles that are apparent only at small cutoffs.  Overall, good agreement is found between the analytical and numerical calculations for the asymptotic behavior of the NLO PV scattering amplitudes.  The slight disagreement at large cutoffs for the $g_{1}$ LEC in the $\DS\!\to\!\QPoh$ channel is a numerical issue occurring at the poles of the NLO three-body force and stemming from numerical fine tuning.  Changing the number of mesh points can noticeably change the appearance of this discrepancy and further methods will need to be developed to deal with it properly.

\begin{figure}[hbt]
\includegraphics[width=100mm]{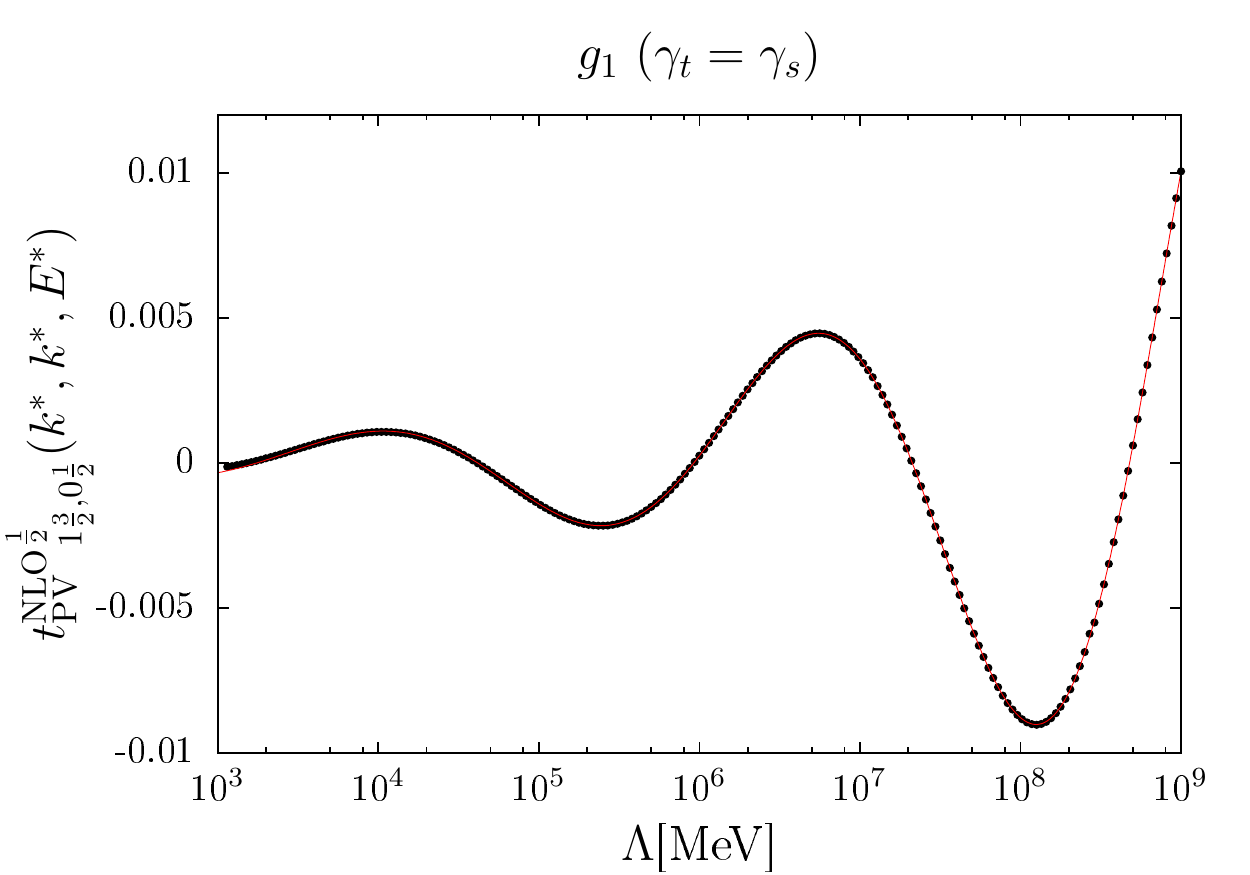}
\caption{\label{fig:NLOAsumptotics-Ws} Asymptotic behavior of the NLO PV scattering amplitude for the $\DS\!\to\!\QPoh$ channel and the LEC $g_{1}$ in the Wigner limit ($\gamma_{t}=\gamma_{s}$).  $k^{*}=1$~MeV and $E^{*}=\frac{3}{4}\frac{(k^{*})^{2}}{M_{N}}-\frac{\gamma_{t}^{2}}{M_{N}}$.}
\end{figure}

Since $D^{\QPoh}$ and $H^{\QPoh}$ are zero for $g_{1}$, $E^{\QPoh}$ is the only contribution to the divergent asymptotic behavior in Eq.~(\ref{eq:PVNLOaympQP}).  However, if $\rho_{t}=\rho_{s}$ then this term is also zero.  Therefore, in the limit $\rho_{t}=\rho_{s}$ there is no divergence for $g_{1}$ in the $\DS\!\to\!\QPoh$ channel and this is observed numerically.  It should also be noted that in the Wigner limit ($\gamma_{t}=\gamma_{s}$) the last two lines of Eq.~(\ref{eq:PVNLOaympQP}) for $g_{1}$ in the $\DS\!\to\QPoh$ channel disappear since the contribution to $D^{\QPoh}_{-}$ from $g_{1}$ comes with a factor of [$\gamma_{t}-\gamma_{s}$] in front.  This means that in the Wigner limit the asymptotic behavior for $g_{1}$ in the $\DS\!\to\!\QPoh$ contains no contribution from $H_{\mathrm{NLO}}(\Lambda)$ and hence has no poles even at small cutoffs as visible in Fig.~\ref{fig:NLOAsumptotics-Ws}.

Overall it is found that the NLO PV scattering amplitude in the $\DS\!\to\!\DPoh$ and $\DS\!\to\!\QPoh$ channels diverges like $\Lambda^{2-s_{1}}$ for large $\Lambda$.  To remove this divergence a NLO PV three-body force will be required.  The LO PC quartet $S$-wave channel goes asymptotically like $q^{-3.16622...}$ instead of $q^{-1}$ as in the LO PC doublet $S$-wave channel~\cite{Griesshammer:2005ga}.  Therefore, NLO PV scattering amplitudes in the channels $\QS\!\to\!\DPoh$ and $\QS\!\to\!\QPoh$ converge as $\Lambda\to\infty$ and this is observed numerically.  All $P$-wave to $D$-wave transitions up to NLO also converge as $\Lambda\to\infty$ and this is observed numerically and can be shown analytically using the asymptotic form of the Ws and Was $D$-wave given in Ref.~\cite{Griesshammer:2005ga}.

\section{\label{sec:observables}Observables}

To calculate PV observables the scattering amplitude $M_{m_{1}',m_{2}';m_{1},m_{2}}$ in the spin basis is related to the scattering amplitude in the total angular momentum basis via 
\begin{align}
\label{eq:relation}
M_{m_{1}',m_{2}';m_{1},m_{2}}=&\sqrt{4\pi}\sum_{J}\sum_{L,L'}\sum_{S,S'}\sum_{m_{S},m_{S}'}\sum_{m_{L'}}\sqrt{2L+1}\CG{1}{\frac{1}{2}}{S}{m_{1}}{m_{2}}{m_{S}}\CG{1}{\frac{1}{2}}{S'}{m_{1}'}{m_{2}'}{m_{S}'}\\\nonumber
&\times\CG{L}{S}{J}{0}{m_{s}}{M}\CG{L'}{S'}{J}{m_{L}'}{m_{S}'}{M}Y_{L'}^{m_{L}'}(\theta,\phi)M^{J}_{L'S',LS},
\end{align}
where $m_{1}$ ($m_{2}$) is the initial spin of the deuteron (nucleon) and $m_{1}'$ ($m_{2}'$) is the final spin of the deuteron (nucleon).  The amplitude $M^{J}_{L'S',LS}$ is related to the numerically calculated scattering amplitudes by 
\begin{equation}
M^{J}_{L'S',LS}=Z_{\mathrm{LO}}t^{J}_{L'S',LS}(k,p,E),
\end{equation}
where $Z_{\mathrm{LO}}$ is the LO deuteron wavefunction renormalization in Eq.~(\ref{eq:renorm}) and $t^{J}_{L'S',LS}(k,p,E)$ can be either PC or PV.  One possible PV experiment is neutron spin rotation through a deuterium target.  In this experiment, the de Broglie wavelength of the neutron must be larger than the average spacing between deuterium atoms so that the neutron interacts coherently.  Thus the neutron must have low energies for which only $S$ and $P$-waves are relevant.  Spin rotation has been calculated previously in \EFT using only $S$ and $P$-waves~\cite{Griesshammer:2011md,Vanasse:2011nd}.  However, there are other PV observables that are preferable to perform at higher energies for which $D$-wave contributions are more relevant. One such observable is the longitudinal asymmetry, $A_{L}^{\vec{N}\!d}$, which is the asymmetry constructed from the cross sections of the deuteron unpolarized and the nucleon polarized along and opposite the scattering axis.  In terms of the scattering amplitude in the spin-basis, $A_{L}^{\vec{N}\!d}$ is given by
\begin{equation}
A_{L}^{\vec{N}\!d}=\frac{\sum_{m_{1}',m_{2}'}\sum_{m_{1},m_{2}}(-1)^{\frac{1}{2}-m_{2}}\int d\Omega |M_{m_{1}',m_{2}';m_{1},m_{2}}|^{2}}{\sum_{m_{1}',m_{2}'}\sum_{m_{1},m_{2}}\int d\Omega |M_{m_{1}',m_{2}';m_{1},m_{2}}|^{2}}.
\end{equation}
Using Eq.~(\ref{eq:relation}), $A_{L}^{\vec{N}\!d}$, in terms of scattering amplitudes in the partial wave basis is
\begin{align}
&A_{L}^{\vec{N}\!d}=\frac{2}{3}\mathrm{Re}\left[\left(M^{\frac{1}{2}}_{0\frac{1}{2},0\frac{1}{2}}+M^{\frac{1}{2}}_{1\frac{1}{2},1\frac{1}{2}}\right)\left(M^{\frac{1}{2}}_{1\frac{1}{2},0\frac{1}{2}}\right)^{*}+2\sqrt{2}\left(M^{\frac{1}{2}}_{0\frac{1}{2},0\frac{1}{2}}+M^{\frac{1}{2}}_{1\frac{3}{2},1\frac{3}{2}}\right)\left(M^{\frac{1}{2}}_{1\frac{3}{2},0\frac{1}{2}}\right)^{*}\right.\\\nonumber
&-4\left(M^{\frac{3}{2}}_{0\frac{3}{2},0\frac{3}{2}}+M^{\frac{1}{2}}_{1\frac{1}{2},1\frac{1}{2}}\right)\left(M^{\frac{3}{2}}_{1\frac{1}{2},0\frac{3}{2}}\right)^{*}-2\sqrt{5}\left(M^{\frac{3}{2}}_{0\frac{3}{2},0\frac{3}{2}}+M^{\frac{3}{2}}_{1\frac{3}{2},1\frac{3}{2}}\right)\left(M^{\frac{3}{2}}_{1\frac{3}{2},0\frac{3}{2}}\right)^{*}\\\nonumber
&+4\sqrt{2}\left(M^{\frac{1}{2}}_{1\frac{1}{2},1\frac{1}{2}}+M^{\frac{1}{2}}_{2\frac{3}{2},2\frac{3}{2}}\right)\left(M^{\frac{1}{2}}_{2\frac{3}{2},1\frac{1}{2}}\right)^{*}+2\left(M^{\frac{1}{2}}_{1\frac{1}{2},1\frac{1}{2}}+M^{\frac{1}{2}}_{2\frac{1}{2},2\frac{1}{2}}\right)\left(M^{\frac{3}{2}}_{2\frac{1}{2},1\frac{1}{2}}\right)^{*}\\\nonumber
&\left.-20\left(M^{\frac{3}{2}}_{1\frac{3}{2},1\frac{3}{2}}+M^{\frac{1}{2}}_{2\frac{3}{2},2\frac{3}{2}}\right)\left(M^{\frac{1}{2}}_{2\frac{3}{2},1\frac{3}{2}}\right)^{*}-8\sqrt{5}\left(M^{\frac{1}{2}}_{1\frac{3}{2},1\frac{3}{2}}+M^{\frac{3}{2}}_{2\frac{1}{2},2\frac{1}{2}}\right)\left(M^{\frac{3}{2}}_{2\frac{1}{2},1\frac{3}{2}}\right)^{*}\,\right]/\\\nonumber
&\left[\left|M^{\frac{1}{2}}_{0\frac{1}{2},0\frac{1}{2}}\right|^{2}+2\left|M^{\frac{3}{2}}_{0\frac{3}{2},0\frac{3}{2}}\right|^{2}+3\left|M^{\frac{1}{2}}_{1\frac{1}{2},1\frac{1}{2}}\right|^{2}+6\left|M^{\frac{1}{2}}_{1\frac{3}{2},1\frac{3}{2}}\right|^{2}+5\left|M^{\frac{3}{2}}_{2\frac{1}{2},2\frac{1}{2}}\right|^{2}+10\left|M^{\frac{1}{2}}_{2\frac{3}{2},2\frac{3}{2}}\right|^{2}\right],
\end{align}
where all partial waves up to and including $D$-waves have been summed over. By using Eqs.~(\ref{eq:Jone})-(\ref{eq:Jsix}) and the fact that the LO PC amplitudes are independent of $J$ this expression is simplified by relating all partial wave channels differing by their $J$-values to the partial wave channel with the smallest possible value of $J$.  

Another possible PV observable is the PV deuteron vector asymmetry, $T_{10}$, which is constructed from the cross sections with the nucleon unpolarized and the deuteron polarized with spin-1 along and opposite the scattering axis. In terms of the spin-basis scattering amplitude $T_{10}$ is
\begin{equation}
T_{10}=\sqrt{\frac{3}{2}}\frac{\sum_{m_{1}',m_{2}'}\sum_{m_{2}}\int d\Omega \left(|M_{m_{1}',m_{2}';1,m_{2}}|^{2}-|M_{m_{1}',m_{2}';-1,m_{2}}|^{2}\right)}{\sum_{m_{1}',m_{2}'}\sum_{m_{1},m_{2}}\int d\Omega |M_{m_{1}',m_{2}';m_{1},m_{2}}|^{2}},
\end{equation}
in the Madison conventions~\cite{darden1971polarization}.  Transforming to the partial wave basis and summing over all partial waves up to and including $D$-waves gives
\begin{align}
&T_{10}=\\\nonumber
&-\sqrt{\frac{2}{3}}\mathrm{Re}\left[2\left(M^{\frac{1}{2}}_{0\frac{1}{2},0\frac{1}{2}}+M^{\frac{1}{2}}_{1\frac{1}{2},1\frac{1}{2}}\right)\left(M^{\frac{1}{2}}_{1\frac{1}{2},0\frac{1}{2}}\right)^{*}+\sqrt{2}\left(M^{\frac{1}{2}}_{0\frac{1}{2},0\frac{1}{2}}+M^{\frac{1}{2}}_{1\frac{3}{2},1\frac{3}{2}}\right)\left(M^{\frac{1}{2}}_{1\frac{3}{2},0\frac{1}{2}}\right)^{*}\right.\\\nonumber
&-2\left(M^{\frac{3}{2}}_{0\frac{3}{2},0\frac{3}{2}}+M^{\frac{1}{2}}_{1\frac{1}{2},1\frac{1}{2}}\right)\left(M^{\frac{3}{2}}_{1\frac{1}{2},0\frac{3}{2}}\right)^{*}+2\sqrt{5}\left(M^{\frac{3}{2}}_{0\frac{3}{2},0\frac{3}{2}}+M^{\frac{3}{2}}_{1\frac{3}{2},1\frac{3}{2}}\right)\left(M^{\frac{3}{2}}_{1\frac{3}{2},0\frac{3}{2}}\right)^{*}\\\nonumber
&+2\sqrt{2}\left(M^{\frac{1}{2}}_{1\frac{1}{2},1\frac{1}{2}}+M^{\frac{1}{2}}_{2\frac{3}{2},2\frac{3}{2}}\right)\left(M^{\frac{1}{2}}_{2\frac{3}{2},1\frac{1}{2}}\right)^{*}+4\left(M^{\frac{1}{2}}_{1\frac{1}{2},1\frac{1}{2}}+M^{\frac{1}{2}}_{2\frac{1}{2},2\frac{1}{2}}\right)\left(M^{\frac{3}{2}}_{2\frac{1}{2},1\frac{1}{2}}\right)^{*}\\\nonumber
&\left.+20\left(M^{\frac{3}{2}}_{1\frac{3}{2},1\frac{3}{2}}+M^{\frac{1}{2}}_{2\frac{3}{2},2\frac{3}{2}}\right)\left(M^{\frac{1}{2}}_{2\frac{3}{2},1\frac{3}{2}}\right)^{*}-4\sqrt{5}\left(M^{\frac{1}{2}}_{1\frac{3}{2},1\frac{3}{2}}+M^{\frac{3}{2}}_{2\frac{1}{2},2\frac{1}{2}}\right)\left(M^{\frac{3}{2}}_{2\frac{1}{2},1\frac{3}{2}}\right)^{*}\,\right]/\\\nonumber
&\left[\left|M^{\frac{1}{2}}_{0\frac{1}{2},0\frac{1}{2}}\right|^{2}+2\left|M^{\frac{3}{2}}_{0\frac{3}{2},0\frac{3}{2}}\right|^{2}+3\left|M^{\frac{1}{2}}_{1\frac{1}{2},1\frac{1}{2}}\right|^{2}+6\left|M^{\frac{1}{2}}_{1\frac{3}{2},1\frac{3}{2}}\right|^{2}+5\left|M^{\frac{3}{2}}_{2\frac{1}{2},2\frac{1}{2}}\right|^{2}+10\left|M^{\frac{1}{2}}_{2\frac{3}{2},2\frac{3}{2}}\right|^{2}\right],
\end{align}
where again use of Eqs.~(\ref{eq:Jone})-(\ref{eq:Jsix}) and the fact that the LO PC amplitudes are independent of $J$ have been used to simplify this expression.

To garner insight into the contributions of various LECs to $A_{L}^{\vec{N}d}$ it is transformed into the large-$N_{C}$ basis of LECs in Eq.~(\ref{eq:LECLargeNc}) giving
\begin{equation}
\label{eq:ALNd}
A_{L}^{\vec{N}d}=g_{1}^{(N_{C})}A_{L}^{(1)}+g_{3}^{(N_{C}^{-1})}A_{L}^{(3)}+\tau_{3}\left(g_{4}^{(N_{C}^{-1})}A_{L}^{(4)}+g_{5}^{(N_{C}^{-1})}A_{L}^{(5)}\right),
\end{equation}
where $\tau_{3}$ is +1 (-1) for $pd$ ($nd$) scattering.  Although Coulomb contributions are not included they give corrections of the size $\alpha M_{N}/p$.  Thus for $pd$ scattering at $E_{\mathrm{lab}}=1$~MeV Coulomb corrections are $\sim$24\%, which is less than the LO error from \EFT power counting, and as the energy increases contributions from Coulomb corrections become even smaller.  Values of $A_{L}^{(n)}$ for various lab energies are given in Table~\ref{tab:ALvalues} and contributions from $A_{L}^{(1)}$, $A_{L}^{(3)}$, and $A_{L}^{(4)}$ are roughly the same size at lower energies and notably larger than $A_{L}^{(5)}$ over all energies.  For higher energies the contribution from $A_{L}^{(1)}$ starts to dominate the other contributions.  However, since $A_{L}^{(1)}$ corresponds to the only LO($\mathcal{O}(N_{C})$) LEC in large-$N_{C}$ counting the longitudinal asymmetry should essentially be determined by the contribution from $A_{L}^{(1)}$.

\begin{table}[hbt]
\begin{tabular}{|c|cccccc|}
\hline
$E_{\mathrm{lab}}$ [MeV] & 1 & 2  & 3 & 5 & 10 & 15 \\\hline
$A_{L}^{(1)}$ [MeV] & \,\,-217.4\,\, & \,\,-296.5\,\, & \,\,-363.1\,\, & \,\,-493.4\,\, & \,\,-752.4\,\, & \,\,-942.6\,\,\\
$A_{L}^{(3)}$ [MeV] & -191.5 & -256.1 & -309.5 & -424.9 & -575.3 & -636.4\\
$A_{L}^{(4)}$ [MeV] & 251.6 & 299.4 & 334.0 & 395.6 & 495.4 & 568.0\\
$A_{L}^{(5)}$ [MeV] & 29.16 & 45.49 & 60.33 & 77.10 & 199.2 & 344.5\\\hline
\end{tabular}
\caption{\label{tab:ALvalues}Coefficients in front of LECs in large-$N_{C}$ basis for nucleon longitudinal asymmetry, $A_{L}^{\vec{N}d}$ (See Eq.~(\ref{eq:ALNd})).}
\end{table}

Transforming the deuteron vector asymmetry $T_{10}$ to the large-$N_{C}$ basis gives
\begin{equation}
\label{eq:T10}
T_{10}=g_{1}^{(N_{C})}T_{10}^{(1)}+g_{3}^{(N_{C}^{-1})}T_{10}^{(3)}+\tau_{3}\left(g_{4}^{(N_{C}^{-1})}T_{10}^{(4)}+g_{5}^{(N_{C}^{-1})}T_{10}^{(5)}\right),
\end{equation}
and the contribution from each LEC is given in Table~\ref{tab:ADvalues}.  The value from $T_{10}^{(4)}$ dominates at low energies and the contribution from $T_{10}^{(5)}$ is suppressed relative to the other contributions at low energy.  Despite, $T_{10}^{(4)}$ being the dominant contribution at low energies it becomes comparable to $T_{10}^{(1)}$ at higher energies and since $g_{1}^{(N_{C})}$ is the only LO($\mathcal{O}(N_{C})$) contribution in the large-$N_{C}$ basis its contribution should be the important contribution for all energies.
\begin{table}[hbt]
\begin{tabular}{|c|cccccc|}
\hline
$E_{\mathrm{lab}}$ [MeV] & 1 & 2  & 3 & 5 & 10 & 15 \\\hline
$T_{10}^{(1)}$ [MeV] & 105.8 & 156.8 & 206.7 & 306.5 & 460.3 & 572.8\\
$T_{10}^{(3)}$ [MeV] & 81.05 & 117.0 & 154.8 & 243.0 & 292.6& 283.0\\
$T_{10}^{(4)}$ [MeV] & \,\,287.1\,\, & \,\,336.9\,\, & \,\,374.1\,\, & \,\,450.7\,\, & \,\,556.0\,\, & \,\,618.6\,\,\\
$T_{10}^{(5)}$ [MeV] & -27.88 & -44.69 & -58.41 & -71.41 & -188.6 & -326.0\\\hline
\end{tabular}
\caption{\label{tab:ADvalues}Coefficient in front of LECs in large-$N_{C}$ basis for deuteron vector asymmetry $T_{10}$ (See Eq.~(\ref{eq:T10})).}
\end{table}

Using the experimental bound for $A_{L}^{\vec{p}d}$~\cite{Nagle:1978vn}	
\begin{equation}
A_{L}^{\vec{p}d}(E_{\mathrm{lab}}=15~\mathrm{MeV})=(-3.5\pm 8.5)\times 10^{-8},
\end{equation}
and noting that it is primarily determined by the LO($\mathcal{O}(N_{C})$) large-$N_{C}$ contribution $g_{1}^{(N_{C})}$ the bound
\begin{equation}
\label{eq:g1NCbound}
g_{1}^{(N_{C})}=(3.7\pm 9.0)\times 10^{-11}~\mathrm{MeV}^{-1},
\end{equation}
is placed.  Combining this bound with the experimental value of $\vec{pp}$ scattering at 13.6~MeV~\cite{Eversheim:1991tg}
\begin{equation}
A_{L}^{\vec{pp}}(E_{\mathrm{lab}}=13.6~\mathrm{MeV})=(-0.93\pm 0.21)\times 10^{-7},
\end{equation}
and the theoretical \EFT prediction in the large-$N_{C}$ basis~\cite{Phillips:2008hn}
\begin{equation}
A_{L}^{\vec{pp}}(13.6~\mathrm{MeV})=\left[602.7 g_{1}^{(N_{C})} + 904.0 g_{2}^{(N_{C})} - 602.8 g_{3}^{(N_{C}^{-1})} + 904.0 g_{5}^{(N_{C}^{-1})}\right]~\mathrm{MeV},
\end{equation}
gives the value
\begin{equation}
\label{eq:g2NCvalue}
g_{2}^{(N_{C})}=(-1.3\pm 0.83)\times 10^{-10}~\mathrm{MeV}^{-1},
\end{equation}
for the remaining LO($\mathcal{O}(N_{C})$) LEC in large-$N_{C}$.  In this analysis all N$^{2}$LO($\mathcal{O}(N_{C}^{-1})$) LECs in the large-$N_{C}$ counting are dropped.  This bound for $g_{1}^{(N_{C})}$ and prediction for $g_{2}^{(N_{C})}$ are also consistent with the bound from the asymmetry of circularly polarized photons in $np\to d\vec{\gamma}$ near threshold~\cite{Knyazkov:1984lzj}
\begin{equation}
P_{\gamma}=(1.8\pm 1.8)\times 10^{-7},
\end{equation}
which in \EFT in the large-$N_{C}$ basis is given by~\cite{Schindler:2009wd}
\begin{equation}
P_{\gamma}(np\to d\vec{\gamma})=\left[-2804 g_{1}^{(N_{C})} - 1283 g_{2}^{(N_{C})} - 3659 g_{3}^{(N_{C}^{-1})}\right]~\mathrm{MeV}.
\end{equation}

Finally, we consider the impact from the recent results of the NPDGamma experiment.  The photon asymmetry $A_\gamma$ in $\vec{n}p\to d\gamma$ has been calculated previously in \EFT~\cite{Schindler:2009wd}, which in the large-$N_{C}$ basis yields
\begin{equation}
A_{\gamma}=222 g_{4}^{(N_{C}^{-1})}~\mathrm{MeV}.
\end{equation}
Matching onto the experimental~\cite{Blyth:2018aon} value in Eq.~(\ref{eq:NPDGamma}) gives the LEC value
\begin{equation}
g_{4}^{(N_{C}^{-1})}=(-1.4\pm 0.63(stat.)\pm 0.09(syst.))\times 10^{-10}~\mathrm{MeV}^{-1}.	
\end{equation}
Comparing this to the bound on $g_{1}^{(N_{C})}$ Eq.~(\ref{eq:g1NCbound}) and the prediction for $g_{2}^{(N_{C})}$ Eq.~(\ref{eq:g2NCvalue}) it is apparent they are of similar size in disagreement with large-$N_{C}$ counting.  Another recent preliminary analysis also shows inconsistencies in the large-$N_{C}$ counting~\cite{Haxton:CIPANP2018}.

\section{\label{sec:conclusion}Conclusion}

Investigating the asymptotic behavior of the LO PV scattering amplitude it was shown that no LO three-body PV force is required in agreement with Grie{\ss}hammer and Schindler~\cite{Griesshammer:2010nd}.  Continuing the asymptotic analysis to NLO it was demonstrated that a NLO PV three-body force will be needed to absorb divergences that scale roughly like $\Lambda^{0.23}$ in the limit $\Lambda\to\infty$, which disagrees with the claims of Grie{\ss}hammer and Schindler~\cite{Griesshammer:2010nd}.  The arguments made by Grie{\ss}hammer and Schindler relied on the use of Fierz rearrangements and the Wigner basis structure of PV three-body forces.  However, Fierz rearrangements carried out separately in the SU(2) spin and isospin space do not preserve Wigner basis structure and Wigner-SU(4) symmetry invalidating their claims.  Therefore, to predict PV observables to LO ($\sim$30\%) only five two-body PV LECs will be needed, but to predict PV observables for nuclear systems with $A\geq 3$ to NLO ($\sim$10\%) will require at least one additional PV three-body force.  Ideally, this PV three-body force should be fit to three-nucleon data and should be addressed once such data becomes available.

Building on the work of Refs.~\cite{Griesshammer:2011md,Vanasse:2011nd} the calculation of PV $N\!d$ scattering was extended to include thee-body $P$ to $D$-wave transitions.  Using these new transitions the longitudinal asymmetry $A_{L}^{\vec{N}\!d}$ and the deuteron vector asymmetry $T_{10}$ were calculated to higher energies than in Ref.~\cite{Vanasse:2011nd}.  Since the $\Delta I=2$ LEC is suppressed in three-nucleon systems only one LO($\mathcal{O}(N_{C})$) in large-$N_{C}$ LEC, $g_{1}^{(N_{C})}$, appears at LO in \EFT. Over a wide range of energies this work finds $A_{L}^{\vec{N}\!d}$ and $T_{10}$ should essentially be dominated by the LEC $g_{1}^{(N_{C})}$ in the large-$N_{C}$ basis.  Using this fact with an experimental bound for $A_{L}^{\vec{p}d}$ at 15~MeV~\cite{Nagle:1978vn} gives the bound $g_{1}^{(N_{C})}=(3.7\pm 9.0)\times 10^{-11}$~MeV$^{-1}$.  Calculations here did not include Coulomb corrections for $pd$ scattering, but at lab energies of 15~MeV Coulomb corrections are roughly a $\sim$6\% effect and can be ignored compared to the LO \EFT error of $\sim$30\%.  Isospin breaking effects in the PC sector should also only contribute a few percent and can be ignored.  With this bound and the experimental measurement of $pp$ scattering at 13.6~MeV~\cite{Eversheim:1991tg} the value for the remaining LO($\mathcal{O}(N_{C})$) LEC in large-$N_{C}$ can be predicted yielding $g_{2}^{(N_{C})}=(-1.3\pm 0.83)\times 10^{-10}$~MeV$^{-1}$.  Fitting to the recent measurement of $A_{\gamma}$~\cite{Blyth:2018aon} for $\vec{n}p\to d\gamma$ from the NPDGamma collaboration gives $g_{4}^{(N_{C}^{-1})}=(-1.4\pm 0.63(stat.)\pm 0.09(syst.))\times 10^{-10}$~MeV$^{-1}$ for a N$^{2}$LO($\mathcal{O}(N_{C}^{-1})$) in large-$N_{C}$ LEC.  The LO($\mathcal{O}(N_{C})$) and N$^{2}$LO($\mathcal{O}(N_{C}^{-1})$) LECs in large-$N_{C}$ appear to be of similar size, in apparent contradiction with the large-$N_{C}$ hierarchy.  

In this large-$N_{C}$ analysis the experimental data is at lab energies of 15~MeV in the $N\!d$ system and 13.6~MeV in the $\NN$ system, which are equivalent to c.m. momenta of 112~MeV and 80~MeV respectively.  These momenta are both less than the naive breakdown scale of \EFT $\Lambda_{\not{\pi}}\sim m_{\pi}$, but are close to it.  A rigorous analysis of errors both theoretical~\cite{Furnstahl:2014xsa,Furnstahl:2015rha} and experimental should be carried out in future and could potentially mitigate some of the observed discrepancy in large-$N_{C}$.  More experiments of few-body systems at lower energies will make an analysis of the relative scaling of LECs clearer.  It should also be noted that the factor of $\sin^{2}\theta_W$ in the $g_{2}^{(N_{C})}$ coefficient is observed to be unimportant at hadronic scales~\cite{Schindler:2015nga}.  If this pattern applies to the isovector LECs then the large-$N_{C}$ scaling of one of the two isovector LECs would be suppressed by only a factor of $1/N_{C}$ vs. the current $1/N_{C}^{2}$ relative to the LO($\mathcal{O}(N_{C})$) in large-$N_{C}$ LECs.  This would also help to mitigate the observed discrepancy in large-$N_{C}$.  	However, this analysis suggests disagreement between experiment and the current large-$N_{C}$ analysis of the five PV two-body LECs in agreement with another preliminary analysis~\cite{Haxton:CIPANP2018}.

\acknowledgments{I would like to thank Roxanne Springer, Matthias Schindler, and Harald Grie{\ss}hammer for useful discussions during the course of this work.  This material is based
upon work supported by the U.S. Department of Energy, Office of Science, Office of Nuclear Physics, under Award Number DE-FG02-05ER41368}
\appendix

\section{\label{app:Projections} Projections}

\subsection{Parity-conserving}

Following the methods of Ref.~\cite{Margaryan:2015rzg} the projection for the LO PC kernel in angular momentum and isospin is
\begin{align}
&\left[{\mathbf{K}_{\mathrm{PC}}}^{J}_{L'S'T',LST}(k,p,E)\right]_{xy}=2\pi\sqrt{\widehat{x}\widehat{y}\widehat{1-x}\widehat{1-y}}\left(4\SJ{x}{\frac{1}{2}}{\frac{1}{2}}{y}{S}{\frac{1}{2}}\SJ{1-x}{\frac{1}{2}}{\frac{1}{2}}{1-y}{T}{\frac{1}{2}}\frac{1}{kp}Q_{L}(a)\right.\\\nonumber
&\hspace{2cm}\left.\vphantom{\SJ{x}{\frac{1}{2}}{\frac{1}{2}}{y}{S}{\frac{1}{2}}}+\frac{4\pi H_{\mathrm{LO}}(\Lambda)}{\Lambda^{2}}\delta_{L0}\delta_{S\nicefrac{1}{2}}\delta_{T\nicefrac{1}{2}}\right)\delta_{LL'}\delta_{SS'}\delta_{TT'},
\end{align}
where $T$ ($T'$) is the total initial (final) isospin of the $N\!d$ system.  The values of $x$ and $y$ pick out the matrix elements in c.c.~space, where $x=1$ ($y=1$) refers to an initial (final) spin-triplet dibaryon propagator and $x=0$ ($y=0$) to an initial (final) spin-singlet dibaryon propagator.  In the Legendre function of the second kind the value of $a$ is given in Eq.~(\ref{eq:a}).

\subsection{Parity-violating}
The LO PV kernel projected out into angular momentum and isospin is split into three parts
\begin{align}
\left[{\mathbf{K}_{\mathrm{PV}}}^{J}_{L'S'T',LST}(k,p,E)\right]_{xy}=\left[{\mathbf{K}_{\mathrm{PV}}^{(\mathrm{\Rmnum{1}})}}^{J}_{L'S'T',LST}(k,p,E)\right]_{xy}&+\left[{\mathbf{K}_{\mathrm{PV}}^{(\mathrm{\Rmnum{2}})}}^{J}_{L'S'T',LST}(k,p,E)\right]_{xy}\\\nonumber
&+\left[{\mathbf{K}_{\mathrm{PV}}^{(\mathrm{\Rmnum{3}})}}^{J}_{L'S'T',LST}(k,p,E)\right]_{xy},
\end{align}
where parts $(\mathrm{\Rmnum{1}})$, $(\mathrm{\Rmnum{2}})$, and $(\mathrm{\Rmnum{3}})$ are the contributions from the two-body PV LECs $g_{1}$, $g_{2}$, and $g_{3}$, $g_{4}$, and $g_{5}$ respectively.  Each part can then be split up into a contribution from each of the tree-level diagrams-(a) and (b) shown in Fig.~\ref{fig:treelevel}
\begin{equation}
\left[{\mathbf{K}_{\mathrm{PV}}^{(\mathrm{X})}}^{J}_{L'S'T',LST}(k,p,E)\right]_{xy}=\left[{\mathbf{K}_{\mathrm{PV}}^{(\mathrm{X})}}^{J}_{L'S'T',LST}(k,p,E)\right]^{(a)}_{xy}+\left[{\mathbf{K}_{\mathrm{PV}}^{(\mathrm{X})}}^{J}_{L'S'T',LST}(k,p,E)\right]^{(b)}_{xy},
\end{equation}
where $\mathrm{X}=\mathrm{\Rmnum{1}}$, $\mathrm{\Rmnum{2}}$, or $\mathrm{\Rmnum{3}}$.  The contribution from diagram (a) and (b) can be related by time reversal symmetry giving
\begin{equation}
\left[{\mathbf{K}_{\mathrm{PV}}^{(\mathrm{X})}}^{J}_{L'S'T',LST}(k,p,E)\right]^{(b)}_{xy}=\left[{\mathbf{K}_{\mathrm{PV}}^{(\mathrm{X})}}^{J}_{LST,L'S'T'}(p,k,E)\right]^{(a)}_{yx}.
\end{equation}
For diagram-(a), the kernel for $\mathrm{X}=\mathrm{\Rmnum{1}}$ is given by
\begin{align}
&\left[{\mathbf{K}_{\mathrm{PV}}^{(\mathrm{\Rmnum{1}})}}^{J}_{L'S'T',LST}(k,p,E)\right]^{(a)}_{xy}=(-1)^{1+y+L+S-J}\sqrt{\widehat{1-x}\widehat{x}\widehat{S'}\widehat{L}}\SJ{S}{y}{S'}{L'}{J}{L}\\\nonumber
&\hspace{3cm}\times\CG{L}{1}{L'}{0}{0}{0}\delta_{TT'}\delta_{1y}\delta_{S\nicefrac{1}{2}}\left[2kQ_{L'}(a)+pQ_{L}(a)\right]g_{1},
\end{align}
for $X=\mathrm{\Rmnum{2}}$ by
\begin{align}
&\left[{\mathbf{K}_{\mathrm{PV}}^{(\mathrm{\Rmnum{2}})}}^{J}_{L'S'T',LST}(k,p,E)\right]^{(a)}_{xy}=(-1)^{S-S'+L'-J-T}12\sqrt{6\widehat{x}\widehat{1-x}\widehat{S}\widehat{S'}\widehat{L'}}\\\nonumber
&\hspace{2cm}\times\SJ{x}{\frac{1}{2}}{\frac{1}{2}}{1}{S}{\frac{1}{2}}\SJ{1}{y}{1}{\frac{1}{2}}{S}{S'}\SJ{S'}{1}{S}{L}{J}{L'}\SJ{1-x}{\frac{1}{2}}{T'}{1}{T}{\frac{1}{2}}\delta_{1y}\delta_{T'\nicefrac{1}{2}}\\\nonumber
&\hspace{2cm}\times\CG{L'}{1}{L}{0}{0}{0}\left[2kQ_{L'}(a)+pQ_{L}(a)\right]g_{2},
\end{align}
and for $\mathrm{X}=\mathrm{\Rmnum{3}}$ by
\begin{align}
&\left[{\mathbf{K}_{\mathrm{PV}}^{(\mathrm{\Rmnum{3}})}}^{J}_{L'S'T',LST}(k,p,E)\right]^{(a)}_{xy}=(-1)^{3/2+L'-S-J}2\sqrt{3\widehat{x}\widehat{S}\widehat{L'}}&\\\nonumber
&\hspace{2cm}\times\SJ{x}{\frac{1}{2}}{S'}{1}{S}{\frac{1}{2}}\SJ{S'}{1}{S}{L}{J}{L'}\CG{L'}{1}{L}{0}{0}{0}\delta_{0y}\delta_{S'\nicefrac{1}{2}}\left[pQ_{L}(a)+2kQ_{L'}(a)\right]\\\nonumber
&\hspace{2cm}\times\left(I^{(0)}_{T',T}(x,y)g_{3}+I^{(1)}_{T',T}(x,y)g_{4}+I^{(2)}_{T',T}(x,y)g_{5}\right),
\end{align}
where the isospin projections are given by
\begin{align}
I^{(0)}_{T',T}(x,y)-2\sqrt{\widehat{1-x}\widehat{1-y}}\SJ{1-y}{\frac{1}{2}}{\frac{1}{2}}{1-x}{T}{\frac{1}{2}}\delta_{TT'},
\end{align}
\begin{equation}
I^{(1)}_{T',T}(x,y)=(-1)^{\frac{1}{2}-y+T'}6\sqrt{2\widehat{1-x}\widehat{T'}}\SJ{1}{\frac{1}{2}}{\frac{1}{2}}{1-x}{T}{\frac{1}{2}}\SJ{\frac{1}{2}}{1-y}{T'}{1}{T}{1}\CG{T}{1}{T'}{m_{T}}{0}{m_{T'}},
\end{equation}
and
\begin{equation}
I^{(2)}_{T',T}(x,y)=(-1)^{\frac{1}{2}-x-y+T'}6\sqrt{10\widehat{1-x}\widehat{T'}}\SJ{1}{\frac{1}{2}}{\frac{1}{2}}{1-x}{T}{\frac{1}{2}}\SJ{\frac{1}{2}}{1-y}{T'}{2}{T}{1}\CG{T}{2}{T'}{m_{T}}{0}{m_{T'}}.
\end{equation}

\section{\label{app:PCasymptotic}Asymptotic analysis of PC amplitudes}

Expanding the LO quartet $P$-wave and the LO Was doublet $P$-wave scattering amplitude in the asymptotic limit ,$q\sim p\gg k,E,\gamma_{t}$, and $\gamma_{s}$, gives
\begin{align}
{t_{\mathrm{PC}}}^{\frac{1}{2}}_{1X,1X}(p)=&\frac{4}{\sqrt{3}\pi}\frac{1}{p}\int_{0}^{\infty}dqQ_{1}\left(\frac{q}{p}+\frac{p}{q}\right){t_{\mathrm{PC}}}^{\frac{1}{2}}_{1X,1X}(q)\\\nonumber
&+\frac{4}{\sqrt{3}\pi}\frac{1}{p}\frac{2}{\sqrt{3}}(\gamma_{t}+\delta_{X\frac{1}{2}}\gamma_{s})\int_{0}^{\infty}dq\frac{1}{q}Q_{1}\left(\frac{q}{p}+\frac{p}{q}\right){t_{\mathrm{PC}}}^{\frac{1}{2}}_{1X,1X}(q),	
\end{align}
where the $\gamma_{t}+\delta_{X\frac{1}{2}}\gamma_{s}$ term comes from expanding the dibaryon propagators, and $X=\frac{1}{2}$ ($X=\frac{3}{2}$) for the $\DPoh$ ($\QPoh$) channel.  Taking the ansatz ${t_{\mathrm{PC}}}^{\frac{1}{2}}_{1X,1X}(p)=B^{^{\hat{X}}\!P_{\nicefrac{1}{2}}}p^{-s_{1}-1}+B_{-1}^{^{\hat{X}}\!P_{\nicefrac{1}{2}}}p^{-s_{1}-2}$ and making the substitution $q=xp$ gives
\begin{align}
B^{^{\hat{X}}\!P_{\nicefrac{1}{2}}}p^{-s_{1}-1}+B^{^{\hat{X}}\!P_{\nicefrac{1}{2}}}_{-1}p^{-s_{1}-2}=&\frac{4}{\sqrt{3}\pi}B^{^{\hat{X}}\!P_{\nicefrac{1}{2}}}p^{-s_{1}-1}\int_{0}^{\infty}dxQ_{1}\left(x+\frac{1}{x}\right)x^{-s_{1}-1}\\\nonumber
&+\frac{4}{\sqrt{3}\pi}B^{^{\hat{X}}\!P_{\nicefrac{1}{2}}}_{-1}p^{-s_{1}-2}\int_{0}^{\infty}dxQ_{1}\left(x+\frac{1}{x}\right)x^{-s_{1}-2}\\\nonumber
&+\frac{4}{\sqrt{3}\pi}B^{^{\hat{X}}\!P_{\nicefrac{1}{2}}}p^{-s_{1}-2}\frac{2}{\sqrt{3}}(\gamma_{t}+\delta_{X\frac{1}{2}}\gamma_{s})\int_{0}^{\infty}dx Q_{1}\left(x+\frac{1}{x}\right)x^{-s_{1}-2}.
\end{align}
The resulting integrals are Mellin transforms defined in Eq.~(\ref{eq:Mellin}), solved in Ref.~\cite{Griesshammer:2005ga}, and give the solution
\begin{align}
B^{^{\hat{X}}\!P_{\nicefrac{1}{2}}}p^{-s_{1}-1}+B^{^{\hat{X}}\!P_{\nicefrac{1}{2}}}_{-1}p^{-s_{1}-2}=&\frac{1}{2}B^{^{\hat{X}}\!P_{\nicefrac{1}{2}}}p^{-s_{1}-1}\mathcal{M}[1,-s_{1}]\\\nonumber
&+\frac{1}{2}B^{^{\hat{X}}\!P_{\nicefrac{1}{2}}}_{-1}p^{-s_{1}-2}\mathcal{M}[1,-s_{1}-1]\\\nonumber
&+B^{^{\hat{X}}\!P_{\nicefrac{1}{2}}}p^{-s_{1}-2}\frac{(\gamma_{t}+\delta_{X\frac{1}{2}}\gamma_{s})}{\sqrt{3}}\mathcal{M}[1,-s_{1}-1].
\end{align}
Equating polynomial coefficients on both sides gives the transcendental equation
\begin{equation}
\frac{1}{2}\mathcal{M}[1,-s_{1}]=1,
\end{equation}
for the value of $s_{1}$, and 
\begin{equation}
B^{^{\hat{X}}\!P_{\nicefrac{1}{2}}}_{-1}=B^{^{\hat{X}}\!P_{\nicefrac{1}{2}}}\frac{\left(\gamma_{t}+\delta_{X\frac{1}{2}}\gamma_{s}\right)}{\sqrt{3}}\frac{\mathcal{M}[1,-s_{1}-1]}{1-\frac{1}{2}\mathcal{M}[1,-s_{1}-1]}.
\end{equation}

\section{\label{app:PVasymptotic} Asymptotic analysis of PV amplitudes}

To ascertain the asymptotic behavior of the LO PV scattering amplitude due to the two-body PV LECs one looks at the inhomogeneous part of Eq.~(\ref{eq:PVLOWig}) which gives
\begin{equation}
\frac{1}{2\pi^{2}}\int_{0}^{\Lambda}dqq^{2}{\mathbf{K_{W}}_{\mathrm{PV}}}^{J}_{L'S',LS}(q,p,E)\mathbf{D_{W}}(E,q){\mathbf{t_{W}}_{\mathrm{PC}}}^{J}_{LS,LS}(k,q,E).
\end{equation}
The leading contribution to the Was part of the $\DS\!\to\!\DPoh$ and $\DS\!\to\!\QPoh$ PV scattering amplitudes comes from the Ws part of the $\DS$ PC scattering amplitude that scales like $Cq^{is_{0}-1}$ in the asymptotic limit.  Expanding the inhomogeneous term and the kernel in the asymptotic limit gives
\begin{align}
{t_{\mathrm{PV}}}^{\frac{1}{2};W\!as}_{1X,0\frac{1}{2}}(p)=\frac{4}{\sqrt{3}\pi }\int_{0}^{\infty}dq\frac{1}{p}\left(f^{(X)}_{W\!as}\,pQ_{0}\left(\frac{q}{p}+\frac{p}{q}\right)-g_{W\!as}^{(X)}\,qQ_{1}\left(\frac{q}{p}+\frac{p}{q}\right)\right)Cq^{is_{0}-1}\\\nonumber
+\frac{4}{\sqrt{3}\pi}\frac{1}{p}\int_{0}^{\infty}dqQ_{1}\left(\frac{q}{p}+\frac{p}{q}\right){t_{\mathrm{PV}}}^{\frac{1}{2};W\!as}_{1X,0,\frac{1}{2}}(q)
\end{align}
where the coefficients $f_{W\!as}^{(X)}$ and $g_{W\!as}^{(X)}$ can be read off from Eqs.~(\ref{eq:Wbasis1}) and (\ref{eq:Wbasis3}) for the $\DS\!\to\!\DPoh$ and $\DS\!\to\!\QPoh$ channels respectively and their values are given in Eq.~(\ref{eq:fandgvalue}).  Making the ansatz ${t_{\mathrm{PV}}}^{\frac{1}{2};W\!as}_{1X,0\frac{1}{2}}(p)=H^{{}^{\hat{X}}\!P_{\nicefrac{1}{2}}}p^{is_{0}}$ and setting $q=xp$ gives
\begin{align}
H^{{}^{\hat{X}}\!P_{\nicefrac{1}{2}}}p^{is_{0}}=\frac{4}{\sqrt{3}\pi}p^{is_{0}}C\int_{0}^{\infty}\!\!dx\left(f_{W\!as}^{(X)}Q_{0}\left(x+\frac{1}{x}\right)x^{is_{0}-1}-g_{W\!as}^{(X)}Q_{1}\left(x+\frac{1}{x}\right)x^{is_{0}}\right)\\\nonumber
+\frac{4}{\sqrt{3}\pi}H^{{}^{\hat{X}}\!P_{\nicefrac{1}{2}}}p^{is_{0}}\int_{0}^{\infty}\!\!dxQ_{1}\left(x+\frac{1}{x}\right)x^{is_{0}}.
\end{align}
Dividing out $p^{is_{0}}$ and using Eqs.~(\ref{eq:I}) and (\ref{eq:Mellin}) and the fact $I(is_{0})=1$ gives
\begin{align}
H^{{}^{\hat{X}}\!P_{\nicefrac{1}{2}}}=\frac{1}{2}C\left\{f_{W\!as}^{(X)}-g_{W\!as}^{(X)}\,\mathcal{M}[1,is_{0}+1]\right\}
+\frac{1}{2}H^{{}^{\hat{X}}\!P_{\nicefrac{1}{2}}}\mathcal{M}[1,is_{0}+1],
\end{align}
and solving for $H^{{}^{\hat{X}}\!P_{\nicefrac{1}{2}}}$ gives the value in Eq.~(\ref{eq:Hvalue}).

The leading contribution to the Was part of the $\DPoh\!\to\!\DS$ PV scattering amplitude comes from the Was part of the $\DPoh$ PC scattering amplitude that scales like $B^{\DPoh}q^{-s_{1}-1}$ in the asymptotic limit, while the leading contribution to the Was part of the $\QPoh\!\to\!\DS$ channel comes from the Ws and Was $\QPoh$ PC scattering amplitude which are the same and scale like $B^{\QPoh}q^{-s_{1}-1}$ in the asymptotic limit.  Expanding the inhomogeneous term and the kernel in the asymptotic limit gives
\begin{align}
{t_{\mathrm{PV}}}^{\frac{1}{2};W\!as}_{0\frac{1}{2},1X}(p)=\frac{4}{\sqrt{3}\pi}\int_{0}^{\infty}dq\frac{1}{p}\left(a_{W\!as}^{(X)}\,qQ_{0}\left(\frac{q}{p}+\frac{p}{q}\right)-b_{W\!as}^{(X)}\,pQ_{1}\left(\frac{q}{p}+\frac{p}{q}\right)\right)B^{{}^{\hat{X}}\!P_{\frac{1}{2}}}q^{-s_{1}-1}\\\nonumber
-\frac{4}{\sqrt{3}\pi}\frac{1}{p}\int_{0}^{\infty}dqQ_{0}\left(\frac{q}{p}+\frac{p}{q}\right){t_{\mathrm{PV}}}^{\frac{1}{2};W\!as}_{0\frac{1}{2},1X}(q),
\end{align}
where $a_{W\!as}^{(X)}$ and $b_{W\!as}^{(X)}$ can be read off from Eqs.~(\ref{eq:Wbasis2}) and (\ref{eq:Wbasis4}) for the $\DPoh\!\to\DS$ and $\QPoh\!\to\!\DS$ channels respectively and are given in Eq.~(\ref{eq:aandbvalueWas}).  Taking the ansatz ${t_{\mathrm{PV}}}^{\frac{1}{2};W\!as}_{0\frac{1}{2},1X}(p)=E^{{}^{\hat{X}}\!P_{\nicefrac{1}{2}}}p^{-s_{1}}$ and setting $q=xp$ gives
\begin{align}
E^{{}^{\hat{X}}\!P_{\nicefrac{1}{2}}}p^{-s_{1}}=\frac{4}{\sqrt{3}\pi}p^{-s_{1}}B^{{}^{\hat{X}}\!P_{\nicefrac{1}{2}}}\int_{0}^{\infty}\!\!dx\left(a_{W\!as}^{(X)}Q_{0}\left(x+\frac{1}{x}\right)x^{-s_{1}}-b_{W\!as}^{(X)}Q_{1}\left(x+\frac{1}{x}\right)x^{-s_{1}-1}\right)\\\nonumber
-\frac{4}{\sqrt{3}\pi}E^{{}^{\hat{X}}\!P_{\nicefrac{1}{2}}}p^{-s_{1}}\int_{0}^{\infty}\!\!dxQ_{0}\left(x+\frac{1}{x}\right)x^{-s_{1}}.
\end{align}
Using Eqs.~(\ref{eq:I}) and (\ref{eq:Mellin}) gives
\begin{align}
E^{{}^{\hat{X}}\!P_{\nicefrac{1}{2}}}=\frac{1}{2}B^{{}^{\hat{X}}\!P_{\nicefrac{1}{2}}}\left\{a_{W\!as}^{(X)}\,I(1-s_{1})-b_{W\!as}^{(X)}\,\mathcal{M}[1,-s_{1}]\right\}
-\frac{1}{2}E^{{}^{\hat{X}}\!P_{\nicefrac{1}{2}}}I(1-s_{1}),
\end{align}
and solving for $E^{{}^{\hat{X}}\!P_{\nicefrac{1}{2}}}$ gives the value in Eq.~(\ref{eq:Evalue}).

The leading contribution to the Ws part of the $\DPoh\!\to\!\DS$ and $\QPoh\!\to\!\DS$ channels is the same as the Was part.  Expanding the inhomogeneous term and the kernel in the asymptotic limit gives

\begin{align}
&{t_{\mathrm{PV}}}^{\frac{1}{2};W\!s}_{0\frac{1}{2},1X}(p)=\frac{4}{\sqrt{3}\pi}\int_{0}^{\infty}dq\frac{1}{p}\left(a_{W\!s}^{(X)}qQ_{0}\left(\frac{q}{p}+\frac{p}{q}\right)-b_{W\!s}^{(X)}pQ_{1}\left(\frac{q}{p}+\frac{p}{q}\right)\right)B^{{}^{\hat{X}}\!P_\frac{1}{2}}q^{-s_{1}-1}\\\nonumber
&\hspace{5mm}+\frac{4}{\sqrt{3}\pi}\frac{2(\gamma_{t}+\bar{\delta}_{X\frac{3}{2}}\gamma_{s})}{\sqrt{3}}\int_{0}^{\infty}dq\frac{1}{p}\left(a_{W\!s}^{(X)}qQ_{0}\left(\frac{q}{p}+\frac{p}{q}\right)-b_{W\!s}^{(X)}pQ_{1}\left(\frac{q}{p}+\frac{p}{q}\right)\right)B^{{}^{\hat{X}}\!P_\frac{1}{2}}q^{-s_{1}-2}\\\nonumber
&\hspace{5mm}+\frac{4}{\sqrt{3}\pi}\int_{0}^{\infty}dq\frac{1}{p}\left(a_{W\!s}^{(X)}qQ_{0}\left(\frac{q}{p}+\frac{p}{q}\right)-b_{W\!s}^{(X)}pQ_{1}\left(\frac{q}{p}+\frac{p}{q}\right)\right)B^{{}^{\hat{X}}\!P_\frac{1}{2}}_{-1}q^{-s_{1}-2}\\\nonumber
&\hspace{5mm}+\frac{8}{\sqrt{3}\pi}\frac{1}{p}\int_{0}^{\infty}dqQ_{0}\left(\frac{q}{p}+\frac{p}{q}\right){t_{\mathrm{PV}}}^{\frac{1}{2};W\!s}_{0\frac{1}{2},1X}(q)\\\nonumber
&\hspace{5mm}+\frac{8}{\sqrt{3}\pi}\frac{2(\gamma_{t}+\gamma_{s})}{\sqrt{3}}\frac{1}{p}\int_{0}^{\infty}dq\frac{1}{q}Q_{0}\left(\frac{q}{p}+\frac{p}{q}\right){t_{\mathrm{PV}}}^{\frac{1}{2};W\!s}_{0\frac{1}{2},1X}(q)\\\nonumber
&\hspace{5mm}+\frac{8}{\sqrt{3}\pi}\frac{2(\gamma_{t}-\gamma_{s})}{\sqrt{3}}\frac{1}{p}\int_{0}^{\infty}dq\frac{1}{q}Q_{0}\left(\frac{q}{p}+\frac{p}{q}\right){t_{\mathrm{PV}}}^{\frac{1}{2};W\!as}_{0\frac{1}{2},1X}(q),
\end{align}
where the subleading behavior from the dibaryon propagators and the $^{\hat{X}}\!P_{\nicefrac{1}{2}}$ PC scattering amplitude are included.  The value $\bar{\delta}_{X\frac{3}{2}}=1-\delta_{X\frac{3}{2}}$.  Making the ansatz ${t_{\mathrm{PV}}}^{\frac{1}{2};W\!s}_{0\frac{1}{2},1X}(p)=D^{{}^{\hat{X}}\!P_{\nicefrac{1}{2}}}p^{-s_{1}}+D_{-}^{{}^{\hat{X}}\!P_{\nicefrac{1}{2}}}p^{-s_{1}-1}$ and setting $q=xp$ gives
\begin{align}
&D^{{}^{\hat{X}}\!P_{\nicefrac{1}{2}}}p^{-s_{1}}+D_{-}^{{}^{\hat{X}}\!P_{\nicefrac{1}{2}}}p^{-s_{1}-1}=\\\nonumber
&\frac{4}{\sqrt{3}\pi M_{N}}p^{-s_{1}}B^{{}^{\hat{X}}\!P_{\nicefrac{1}{2}}}\int_{0}^{\infty}\!\!dx\left(a_{W\!s}^{(X)}Q_{0}\left(x+\frac{1}{x}\right)x^{-s_{1}}-b_{W\!s}^{(X)}Q_{1}\left(x+\frac{1}{x}\right)x^{-s_{1}-1}\right)\\\nonumber
&+p^{-s_{1}-1}B^{{}^{\hat{X}}\!P_{\nicefrac{1}{2}}}\frac{8(\gamma_{t}+\bar{\delta}_{X\frac{3}{2}}\gamma_{s})}{3\pi M_{N}}\int_{0}^{\infty}\!\!dx\left(a_{W\!s}^{(X)}Q_{0}\left(x+\frac{1}{x}\right)x^{-s_{1}-1}-b_{W\!s}^{(X)}Q_{1}\left(x+\frac{1}{x}\right)x^{-s_{1}-2}\right)\\\nonumber
&+\frac{4}{\sqrt{3}\pi M_{N}}p^{-s_{1}-1}B^{{}^{\hat{X}}\!P_{\nicefrac{1}{2}}}_{-1}\int_{0}^{\infty}\!\!dx\left(a_{W\!s}^{(X)}Q_{0}\left(x+\frac{1}{x}\right)x^{-s_{1}-1}-b_{W\!s}^{(X)}Q_{1}\left(x+\frac{1}{x}\right)x^{-s_{1}-2}\right)\\\nonumber
&+\frac{8}{\sqrt{3}\pi}D^{{}^{\hat{X}}\!P_{\nicefrac{1}{2}}}p^{-s_{1}}\int_{0}^{\infty}\!\!dxQ_{0}\left(x+\frac{1}{x}\right)x^{-s_{1}}\\\nonumber
&+\frac{8}{\sqrt{3}\pi}D^{{}^{\hat{X}}\!P_{\nicefrac{1}{2}}}_{-1}p^{-s_{1}-1}\int_{0}^{\infty}\!\!dxQ_{0}\left(x+\frac{1}{x}\right)x^{-s_{1}-1}\\\nonumber
&+\frac{8}{\sqrt{3}\pi}\frac{2(\gamma_{t}+\gamma_{s})}{\sqrt{3}}D^{{}^{\hat{X}}\!P_{\nicefrac{1}{2}}}p^{-s_{1}-1}\int_{0}^{\infty}\!\!dxQ_{0}\left(x+\frac{1}{x}\right)x^{-s_{1}-1}\\\nonumber
&+\frac{8}{\sqrt{3}\pi}\frac{2(\gamma_{t}-\gamma_{s})}{\sqrt{3}}E^{{}^{\hat{X}}\!P_{\nicefrac{1}{2}}}p^{-s_{1}-1}\int_{0}^{\infty}\!\!dxQ_{0}\left(x+\frac{1}{x}\right)x^{-s_{1}-1}.
\end{align}
Finally, using Eqs.~(\ref{eq:I}) and (\ref{eq:Mellin}) and collecting the coefficients of polynomial terms gives
\begin{align}
D^{{}^{\hat{X}}\!P_{\nicefrac{1}{2}}}=\frac{1}{2}B^{{}^{\hat{X}}\!P_{\nicefrac{1}{2}}}\left\{a_{W\!s}^{(X)}I(1-s_{1})-b_{W\!s}^{(X)}\mathcal{M}[1,-s_{1}]\right\}
+D^{{}^{\hat{X}}\!P_{\nicefrac{1}{2}}}I(1-s_{1}),
\end{align}
and
\begin{align}
&D^{{}^{\hat{X}}\!P_{\nicefrac{1}{2}}}_{-1}=\frac{1}{2}\left(\frac{2(\gamma_{t}+\bar{\delta}_{X\frac{3}{2}}\gamma_{s})}{\sqrt{3}}B^{{}^{\hat{X}}\!P_{\nicefrac{1}{2}}}+B^{{}^{\hat{X}}\!P_{\nicefrac{1}{2}}}_{-1}\right)\left\{a_{W\!s}^{(X)}I(-s_{1})-b_{W\!s}^{(X)}\mathcal{M}[1,-s_{1}-1]\right\}\\\nonumber
&\hspace{1cm}+D^{{}^{\hat{X}}\!P_{\nicefrac{1}{2}}}\frac{2(\gamma_{t}+\gamma_{s})}{\sqrt{3}}I(-s_{1})+E^{{}^{\hat{X}}\!P_{\nicefrac{1}{2}}}\frac{2(\gamma_{t}-\gamma_{s})}{\sqrt{3}}I(-s_{1})+D^{{}^{\hat{X}}\!P_{\nicefrac{1}{2}}}_{-1}I(-s_{1}),
\end{align}
giving the solutions for $D^{{}^{\hat{X}}\!P_{\nicefrac{1}{2}}}$ and $D_{-}^{{}^{\hat{X}}\!P_{\nicefrac{1}{2}}}$ in Eqs.~(\ref{eq:Dvalue}) and (\ref{eq:Dmvalue}) respectively.


\end{document}